\documentclass[reprint,aps,prb,twocolumn,superscriptaddress,amsmath,amssymb,longbibliography]{revtex4-2}
\usepackage{graphicx}
\usepackage[colorlinks=true,citecolor=blue,
            linkcolor=blue,urlcolor=blue,bookmarks=true]{hyperref}
\usepackage[utf8]{inputenc}
\usepackage[english]{babel}
\usepackage{amsmath}
\usepackage{mathbbol}
\usepackage{multirow}
\usepackage{physics}
\usepackage{booktabs}
\makeatletter
\renewcommand{\fnum@figure}[1]{\textbf{FIG.~\thefigure.~}}
\makeatother

\begin{document}
\raggedbottom 

\title{Simulated Bifurcation Quantum Annealing}

\author{J. Paw\l{o}wski}
\affiliation{Institute of Theoretical Physics, Faculty of Fundamental Problems of Technology, Wroc\l{a}w University of Science and Technology, 50-370 Wroc\l{a}w, Poland}
\affiliation{Quantumz.io Sp.\;z\;o.o., Puławska 12/3, 02-566 Warsaw}

\author{P. Tarasiuk}
\affiliation{Quantumz.io Sp.\;z\;o.o., Puławska 12/3, 02-566 Warsaw}

\author{J. Tuziemski}
\affiliation{Quantumz.io Sp.\;z\;o.o., Puławska 12/3, 02-566 Warsaw}

\author{Ł. Pawela}
\affiliation{Institute of Theoretical and Applied Informatics, Polish Academy of Sciences, Ba{\l}tycka 5, 44-100 Gliwice, Poland}
\affiliation{Quantumz.io Sp.\;z\;o.o., Puławska 12/3, 02-566 Warsaw}

\author{B. Gardas}
\affiliation{Institute of Theoretical and Applied Informatics, Polish Academy of Sciences, Ba{\l}tycka 5, 44-100 Gliwice, Poland}

\begin{abstract}

We introduce Simulated Bifurcation Quantum Annealing (SBQA), a quantum-inspired
optimization algorithm that extends simulated bifurcation by incorporating
inter-replica interactions to mimic quantum tunneling. SBQA retains the
efficiency and parallelism of simulated bifurcation while improving performance
on sparse and rugged energy landscapes. We derive its equations of motion,
analyze parameter dependence, and propose a lightweight auto-tuning strategy. A
comprehensive benchmarking study on both large-scale problems and smaller
instances relevant for current quantum hardware shows that SBQA systematically
improves on SBM in the sparse and rugged regimes where SBM is known to struggle,
while remaining competitive and versatile across a diverse set of tested problem
families. These results position SBQA as a practical quantum-inspired
optimization heuristic and a stronger classical baseline for the sparse and
rugged regimes studied here.

\end{abstract}

\maketitle

\section{Introduction}
Combinatorial optimization problems are ubiquitous in science and industry~\cite{Weinand2022, Martins2025,Rahman2021, Mohseni2022} and 
can often be formulated as the task of finding the ground state, or low-energy states, of an Ising spin-glass 
Hamiltonian~\cite{Lucas2014}. Quantum annealing (QA) offers a promising hardware-based approach to tackling such problems by 
leveraging quantum superposition and tunneling effects~\cite{Yarkoni2022, Yulianti2022, Abbas2024}. However, practical implementations 
remain constrained by limited qubit connectivity~\cite{Pelofske2025,Gomeztejedor2025}, noise, and finite coherence 
times~\cite{King2022, Pelofske2023}, which have so far prevented an unambiguous demonstration of 
quantum speedup or general computational advantage~\cite{DWaveSupremacy,DWaveSupremacyComment1,DWaveSupremacyComment2,Tuziemski2025,Lidar2025,Pawlowski2025}.

In parallel, a rapidly growing class of physics-inspired classical algorithms has emerged, offering competitive performance on standard hardware by
emulating physical processes, including aspects of quantum dynamics \mbox{\cite{Schuetz2022,Honari2022,Zhang2022,Zeng2024}}. Among these approaches,
the Simulated Bifurcation Machine (SBM) stands out as a highly efficient and scalable optimization method
based on nonlinear Hamiltonian dynamics exhibiting chaotic behavior and bifurcations. This dynamical system can be interpreted as a~mean-field
approximation of a network of Kerr parametric oscillators~\cite{Goto2016,Goto2019,Goto2021,Kanao2022}. Despite its efficiency and versatility, 
SBM is known to exhibit a systematic weakness on certain classes of energy landscapes, particularly those featuring steep, isolated optima or very sparse
connectivity~\cite{Katzgraber2025,veloxq}.

In this work, we introduce \emph{Simulated Bifurcation Quantum Annealing} (SBQA), an optimization algorithm designed to address these limitations
by combining the efficiency of SBM with key ingredients of Discrete-Time Simulated Quantum Annealing (DTSQA). DTSQA is derived from a
path-integral Monte Carlo formulation of the transverse-field Ising model at finite temperature~\cite{Chowdhury2025,Camsari2019}.
Through a Suzuki--Trotter decomposition of the quantum partition function, an additional (imaginary-time) dimension is introduced,
mapping the original $d$-dimensional quantum system onto a $(d+1)$-dimensional classical system~\cite{Suzuki1976PIMC}. In this
representation, quantum fluctuations induced by the transverse field appear as effective ferromagnetic couplings between replicas
(imaginary-time slices), while the annealing process is realized by gradually reducing the transverse-field strength.

Motivated by this replica-based picture, we extend the original SBM formulation by introducing an effective inter-replica coupling
between otherwise independent trajectories. This interaction acts as a classical surrogate for quantum tunneling and helps the
dynamics escape local minima. Crucially, even modest performance improvements -- on the order of a few to several percent -- can be
decisive in practice. Recent evidence indicates that claims of quantum advantage often hinge on such narrow margins, with the
distinction between observing an advantage and failing to do so resting on quantitatively small effects. In this sense, improvements
of the magnitude considered here are not merely incremental, but can qualitatively alter the outcome of quantum--classical
performance comparisons~\cite{Pawlowski2025}.

\begin{figure*}[htbp]
	\centering
	\includegraphics[width=0.9\linewidth]{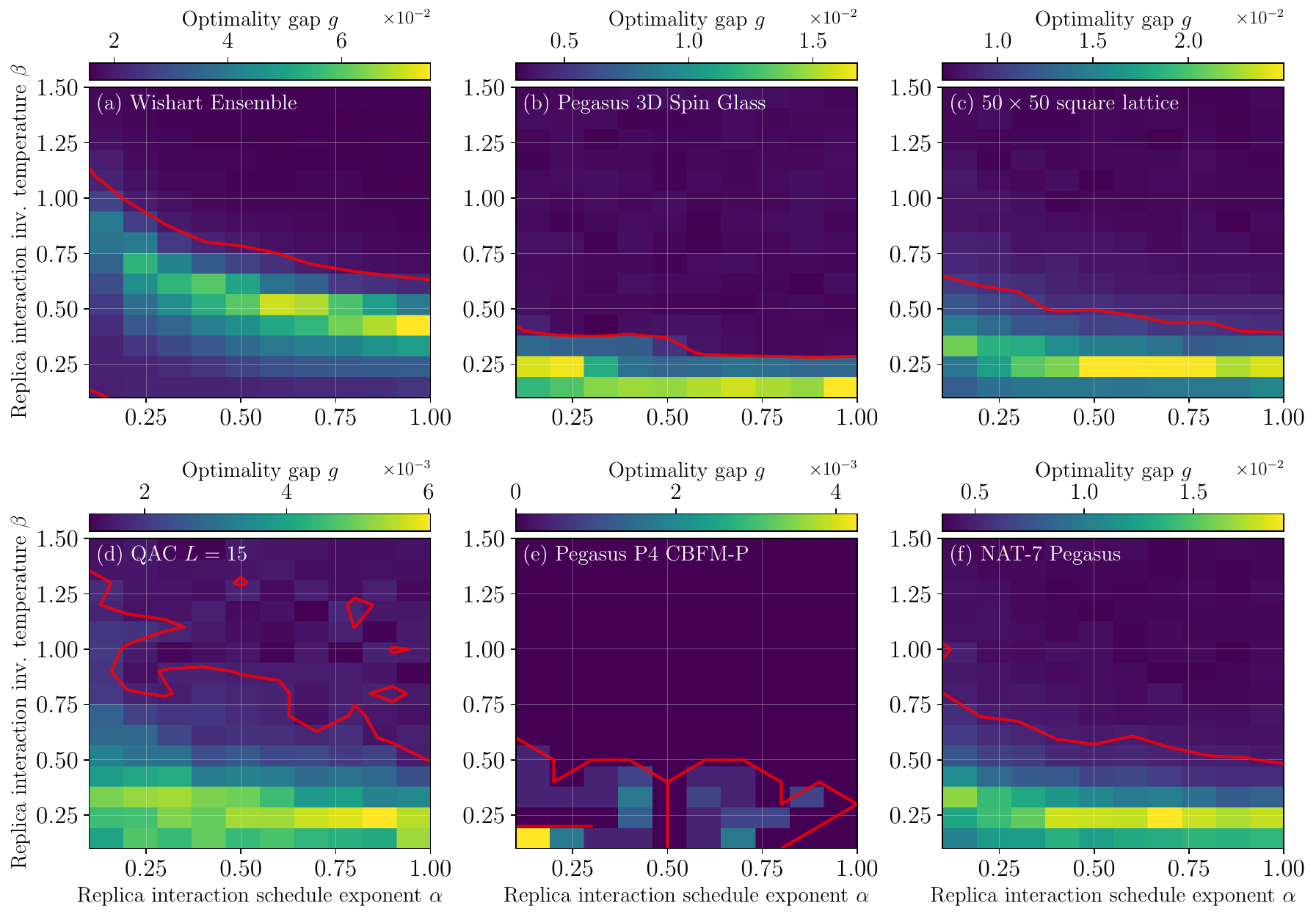}
	\caption{Heatmaps of optimality gaps \(g\) in the \(\alpha, \beta\) plane, demonstrating the
		sensitivity of the SBQA algorithm. Panels (a)--(f) correspond to different instances; see main text for details.
		Plotted values of \(g\) are averaged over best energies from 10 independent runs for each instance, with \(8\)
			sets of \(128\) interacting replicas per run. Red contour lines correspond to a \(35\%\) threshold relative to the best value obtained.
			The results are consistent with interpreting \(\beta\) as the inverse temperature, since energies decrease with
			increasing \(\beta\). They suggest selecting \(\beta\) from the range \(0.5 \lesssim \beta \lesssim 1.5\).
			The impact of \(\alpha\) is less pronounced, but still visible, especially for the all-to-all instance in panel (a).
			We therefore restrict its range to \(0.5 \lesssim \alpha \lesssim 1.0\). Altogether, this analysis suggests that the SBQA algorithm
			is stable with respect to the choice of hyperparameters, and no complicated, instance-dependent tuning is required.
		}

	\label{fig:param_sensitivity}
\end{figure*}

Our contribution is threefold. First, we derive the modified SBM equations of motion with replica interactions and show that this
enhancement has only a minimal performance overhead~\cite{veloxq2025}. Second, we analyze the role of the additional hyperparameters
and propose a~lightweight auto-tuning strategy that avoids expensive instance-specific optimization.

Third, and most importantly, we perform a comprehensive benchmarking study
divided into two parts. Part one focuses on large-scale sparse problems, where
we compare SBM and SBQA in terms of asymptotic time-to-epsilon scaling. We study
Zephyr graphs, which underlie the latest-generation Advantage2 quantum
annealers, instances defined on the logical graphs of the Quantum Annealing
Correction (QAC) error-correction scheme, recently used in an attempt to
demonstrate quantum scaling advantage~\cite{Lidar2025,Pawlowski2025}, and
tile-planting instances defined on square and cubic
lattices~\cite{Perera2020,Hamze2018}. Across these benchmarks, SBQA consistently
improves on SBM, with gains that typically become more pronounced as problem
size increases and density decreases.

Part two concerns problems relevant for current quantum hardware, where we
compare the two algorithms in terms of absolute time-to-epsilon values. This
includes three-dimensional spin glasses embedded into the Pegasus graph of the
D-Wave quantum annealer~\cite{King2023,Chowdhury2025}, and higher-order binary
optimization problems defined on IBM's heavy-hex topology with $N=156$
qubits~\cite{Chandarana2025}. In these benchmarks, we show that SBQA improves on
SBM in the more challenging sparse and rugged settings while maintaining useful
performance across very different instance topologies, where more specialized
solvers, such as DTSQA, can exhibit degraded performance. Taken together, these
results make SBQA a useful classical reference point for future
quantum-classical comparisons on the sparse and rugged benchmark families
studied here.

Throughout our benchmarks, we follow best practices ensuring a fair comparison, including real runtime measurements and averaging over ensembles of random 
instances~\cite{Tuziemski2025}. This avoids common pitfalls such as reliance on runtime proxies, neglect of overheads, and non-representative instance selection.

For the sake of reproducibility, all instances used throughout this work, as well as the best known energies, are publicly available in the companion
data repository~\cite{repository_sbqa}.

The paper is organized as follows. In Section~\ref{sec:methods} we derive the SBQA equations of motion (\ref{sec:derivation})
and analyze their sensitivity to hyperparameters, including practical tuning strategies (\ref{sec:param_sensitivity}). Section~\ref{sec:benchmarks} presents
the benchmarking results, split into asymptotic time-to-epsilon scaling (\ref{sec:scaling}) and
hardware-relevant instances (\ref{sec:hardware}). Finally, Section~\ref{sec:conclusions} concludes with a summary and outlook.

\section{Methods \label{sec:methods}}
\subsection{Simulated Bifurcation Quantum Annealing \label{sec:derivation}}

Our starting point is a classical Hamiltonian describing a system of $N$ particles, interacting via Ising-like interactions,
and subject to a time-dependent external potential:
\begin{align}
	\mathcal{H} = & \,\,\frac{a_0}{2} \sum_{i=1}^{N} p_i^2 + V,      \\
	V =           & \sum_{i=1}^{N} \bigg[ \frac{a_0 - a(t)}{2} q_i^2
		- c_0 h_i q_i - \frac{c_0}{2}\sum_{j=1}^N J_{ij} q_i f(q_j) \bigg],
\end{align}
where \(a(t) = t/T\) and \(T\) is the total time of the evolution.
The trajectories of the particles are confined to an $N$-dimensional unit hypercube, which is achieved by imposing that \(V = \infty\) if any \(q_i > 1\).
Using Hamilton's equations, we can derive the following dynamical system, called the Simulated Bifurcation Machine (SBM):
\begin{equation}
\begin{split}
	\dot{q}_i & = a_0 p_i, \\
	\dot{p}_i & = -\left[a_0 - a(t)\right] q_i + c_0 \bigg(\sum_{j=1}^{N} J_{ij} f(q_j) + h_i\bigg).
\end{split}
\label{eq:sbm}
\end{equation}
with the condition that if \(\abs{q_i} > 1\), then \(q_i \to \mathrm{sign}(q_i)\) and \(p_i \to 0\). These equations are
chaotic and exhibit bifurcations as the time-dependent parameter \(a(t)\) is varied during the evolution.
The system is initialized in a random state and explores the energy landscape, which eventually takes the shape
determined by the couplings \(J_{ij}\) and the external field \(h_i\), encoding the optimization problem, whose solution
can be extracted by taking \(s_i = \mathrm{sign}(q_i)\). Multiple replicas of the system, with different initial
conditions, can be evolved in parallel, allowing for a more thorough exploration of the phase space, and in turn improving
the chances of finding the low-energy local minimum, or even the global minimum of the problem.
This formulation of SBM corresponds to the definitions from Ref.~\cite{Goto2021}, encompassing the two variants, 
bSB with \(f(x) = x\), and dSB with \(f(x) = \mathrm{sign}(x)\). The latter introduces a violation of energy
conservation, which helps the system to escape local minima, in a conceptually similar fashion to quantum tunneling.

Our modification of SBM, which we call the Simulated Bifurcation Quantum Annealing (SBQA), is inspired by the
Simulated Quantum Annealing (SQA), also known as Path Integral Monte Carlo~\cite{Suzuki1976PIMC}. The idea is to couple
the previously independent replicas of the system, via a certain time-dependent interaction derived from the
Suzuki-Trotter decomposition of the quantum partition function of the transverse-field Ising model, which describes the
quantum annealing process. For a composite system of \(N\) particles and \(R\) replicas, the SBQA Hamiltonian reads:
\vspace{-0.5\baselineskip}
\begin{widetext}
	\begin{align}
		\mathcal{H} & = \sum_{i=1}^{N} \sum_{k=1}^{R} \bigg[\frac{a_0}{2R}  p_{i,k}^2 + \frac{a_0 - a(t)}{2R} q_{i,k}^2  - \frac{c_0}{R}
		\bigg(h_i q_{i,k} + \frac{1}{2}\sum_{j=1}^N J_{ij} q_{i,k} q_{j,k}\bigg) - J_{\perp}(t)  q_{i,k} q_{i,k+1} \bigg]                                                                 \\
		J_\perp(t)  & = -\frac{1}{2\beta} \ln \tanh \bigg( \frac{\beta \Gamma_x(t)}{R} \bigg), \quad
		\Gamma_x(t)  = \Gamma_x(0)\left[\left( 1 - t/T \right)^{\alpha} + 10^{-5}\right],
	\end{align}
\end{widetext}
\vspace{-0.5\baselineskip}
where \(J_{\perp}\) is the inter-replica coupling strength (see
Sec.~\ref{sec:sqa_derivation} in Ref.~\cite{supmat} for derivation),
\(\Gamma_x(0)\) sets the initial transverse-field scale, $\beta$ is the
inverse temperature, and \(\alpha\) is the annealing schedule exponent. The
additive \(10^{-5}\) term is a numerical regularization used in the
implementation to prevent the argument of $\tanh$ from vanishing at the final
time step; without it, $J_{\perp}(t)$ becomes singular as $t \to T$. Writing
down Hamilton's equations, we obtain the equations of motion for the SBQA
dynamical system:
\begin{align}
	\dot{q}_{i,k} & =  \frac{a_0}{R} p_{i,k},                                                                                                 \\
	\dot{p}_{i,k} & =  -\bigg[\frac{a_0 - a(t)}{R}\bigg] q_{i,k} + \frac{c_0}{R} \bigg(\sum_{j=1}^{N} J_{ij} f(q_{j,k}) + h_i\bigg) \nonumber \\  &+ J_{\perp} \bigg( q_{i,k-1} + q_{i,k+1} \bigg)
\end{align}
In our implementation, we choose a different form of the function \(f\), introduced in Ref.~\cite{Han2023}, and further 
explored in the context of SBM in Ref.~\cite{Pawlowski2025}. We use the following piecewise function:
\begin{equation}
	f(x) = \begin{cases}
		0                & |x| \leq \Delta(t), \\
		\mathrm{sign}(x) & |x| > \Delta(t),    \\
	\end{cases}
\end{equation}
where \(\Delta(t) = 0.7\frac{t}{T}\) is a time-dependent threshold. Such a choice mitigates discretization errors.

\subsection{Parameter sensitivity and auto-tuning \label{sec:param_sensitivity}}
We now discuss the impact of the two new hyperparameters introduced in SBQA, namely the inverse temperature \(\beta\) and the
annealing schedule exponent \(\alpha\). The former corresponds to the inverse temperature in the quantum partition function,
while the latter governs the ramp-up of the replica interaction strength \(J_{\perp}(t)\). Although \(\alpha=1\) is the
standard choice in quantum annealing, we nevertheless allow it to vary and study its impact. The remaining parameters \(a_0\), \(c_0\),
and \(T\), as well as their tuning procedures, are inherited from the original SBM and retain their interpretation. To avoid
overfitting to a single problem family, we selected a diverse set of instances of various sizes, including both dense and sparse
problems, while maintaining a quantum-annealing-oriented focus. The chosen instances are:
\begin{itemize}
	\item[(a)] fully connected Wishart Ensemble instance with \(N=500\) variables and hardness parameter \(\alpha_W =0.2\)~\cite{Hamze2020},
	\item[(b)] cubic lattice Ising spin glass embedded into the Pegasus \(P_{16}\) graph with \(N\simeq 5400\) variables~\cite{King2023,Chowdhury2025},
	\item[(c)] \(50 \times 50\) square lattice with random uniform couplings drawn from \([-1,1]\),
	\item[(d)] Quantum Annealing Correction (QAC) instance with side length \(L=15\) and \(N = 1322\) variables~\cite{Lidar2025},
	\item[(e)] Pegasus \(P_4\) graph with the Corrupted Bias Ferromagnetic coupling distribution and \(N=216\) variables~\cite{Tasseff2024},
	\item[(f)] Pegasus \(P_{16}\) graph with NAT-7 random couplings~\cite{Schulz2025}.
\end{itemize}

As the figure of merit, we use the optimality gap, defined as:
\begin{equation}
	g = \frac{E - E_{\mathrm{0}}}{\abs{E_0}},
\end{equation}
where \(E_0\) is the best known energy (in particular the ground state energy, if known) and \(E\) is the energy of the returned solution.
We plot the results in Fig.~\ref{fig:param_sensitivity}, where each point is averaged 
over \(10\) independent runs, with \(8\) sets of \(128\) interacting replicas per run. Based on these results, we restrict 
the range of hyperparameters to \(\beta \in [0.5, 1.5]\) and \(\alpha \in [0.5, 1.0]\). Since it is neither obvious how to 
select the optimal values within these ranges, nor desirable to perform costly, per-instance tuning, we introduce a simple 
auto-tuning procedure. The total number of processed samples \(N_{\rm samples}\) is split into \(N_{\rm repetitions}\) with 
interacting \(N_{\rm replicas}\) replicas each. For each repetition, we randomly select \(\beta\) and \(\alpha\) from the 
above ranges, and evolve the system. Finally, we return the best solution from all repetitions. This procedure is simple, 
yet effective, as it allows one to explore a variety of hyperparameter combinations, without incurring significant overhead,
since the parallel capabilities of modern GPUs allow all \(N_{\rm repetitions}\) repetitions to be executed in parallel.

\section{Results \label{sec:benchmarks}}

A typical measure of performance for heuristic optimization algorithms is the time-to-solution (TTS) metric, defined
as the time required to find the optimal solution with a specified probability, e.g. \(99\%\)~\cite{Troyer2014}.
However, in the context of large-scale combinatorial optimization, insisting on finding the optimal solution is often impractical
due to the exponentially growing complexity of the energy landscape. In many real-world applications, the priority shifts from exactness
to utility, and one must often settle for approximate solutions. Therefore, the critical figure of merit is not necessarily the ability
to find the best possible state, but to find high-quality solutions (with some finite tolerance) within a reasonable computational time.

To rigorously quantify this performance, we follow recent works on this topic and adopt the time-to-epsilon metric~\cite{Lidar2025,Pawlowski2025}.
This metric is defined as the expected computational time required to find a solution with an optimality gap below a specific threshold, with a 
specified probability, usually \(p_{\mathrm{target}} = 0.99\),
\begin{equation}
    \mathrm{TT}\varepsilon = \tau \frac{\log(1 - p_{\mathrm{target}})}{\log(1 - p_{\mathrm{success}})},
\end{equation}
where \(\tau\) is the average time per run of the algorithm, and \(p_{\mathrm{success}}\) is the probability of finding a
solution within \(\varepsilon\) of the ground state (or a suitable reference solution) at least once in a single run.

For this metric to be meaningful, one has to be careful with time measurements, ensuring that they reflect the actual, 
externally measurable computational cost of the algorithm. A common pitfall in the case of quantum annealers is to use the annealing
time of a single run, instead of the total operation time, which includes annealing of multiple shots, programming, thermalization, readout, etc.,
and is accessible e.g. as the ``QPU access time'' on D-Wave machines. This is especially important when comparing quantum and classical methods,
to avoid misleading conclusions about runtime supremacy, see, e.g., Ref.~\cite{Tuziemski2025}.

Time-to-epsilon can be studied from two perspectives: as a measure of asymptotic scaling with problem size, or as a practical performance
metric for finite-size problems. Here, we adopt both points of view. For the former,
we study the asymptotic scaling of time-to-epsilon for four classes of large problems: Zephyr graphs with random couplings and fields~\cite{veloxq2025},
tile-planting instances defined on square and cubic lattices~\cite{Perera2020, Hamze2018}, and Sidon28 instances defined on Quantum Annealing Correction
logical graphs~\cite{Lidar2025,Pawlowski2025}.

For the latter, we focus on instances that are directly relevant for current generation quantum hardware: 3D spin glasses embedded into the
Pegasus graph of D-Wave's quantum annealer~\cite{King2023,Chowdhury2025}, and higher-order binary optimization problems defined on IBM's heavy
hex topology with $N=156$ qubits~\cite{Chandarana2025}. These results demonstrate that SBQA is both a practically useful optimization heuristic
for finite-size problems, and, in the spirit of Ref.~\cite{Gangat2026}, a tool for sharpening the threshold for quantum advantage in the asymptotic regime of large problem sizes.

Since all considered instances are random, for each considered algorithm we compute
the median \([\mathrm{TT}\varepsilon]_{\rm med}\) over an ensemble of instances, which is then minimized over the relevant algorithm hyperparameters,
typically the number of steps (classical algorithms) or the annealing time (quantum annealing).

\subsection{Asymptotic scaling of time-to-epsilon \label{sec:scaling}}

We begin with the analysis of scaling properties of time-to-epsilon as a function of problem size. The computational efficiency
and parallelizability of algorithms based on dynamical systems, such as SBM and SBQA, allow us to study much larger problem
sizes than typically considered, on the order of (and exceeding) \(N=10^5\) variables, which is well beyond the capabilities of
current-generation quantum devices. While it still requires fitting and extrapolation to make claims about asymptotic scaling, 
such large sizes significantly reduce sensitivity to finite-size effects and allow for more robust extraction of the scaling exponent.

We note that $\mathrm{TT}\varepsilon$ is a problematic metric to study, as it requires great care and significant computational
resources to obtain reliable results, which is crucial when it is used to compare quantum and classical methods, and 
assess the quantum advantage. This is not our aim here, and we do not attempt to make any advantage claims, but rather to quantitatively
compare SBQA against SBM on similar grounds. Because of that, we slightly relax the analysis, optimizing the $[\mathrm{TT}\varepsilon]_{\rm med}$ 
over number of steps only, keeping the number of trajectories fixed for a given instance class.

\subsubsection{Large-scale Zephyr instances}
\begin{figure}[t]
    \centering
    \includegraphics[width=\linewidth]{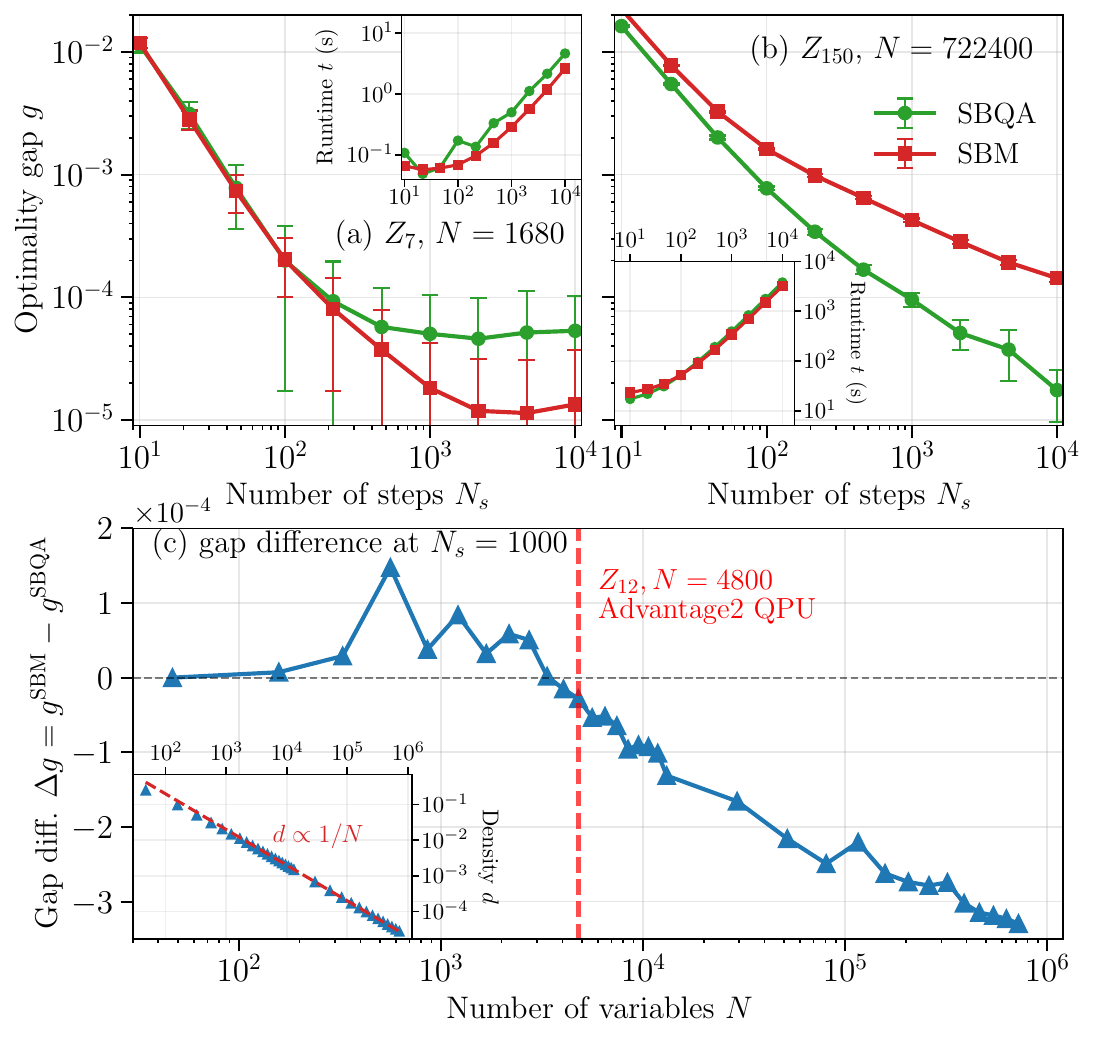}
    \caption{Average optimality gap $g$ as a function of the number of steps for instances on (a) the $Z_7$ graph and (b) the $Z_{150}$ graph, with
    $N=1680$ and $N=722400$ variables, respectively. Results are averaged over $20$ ($5$) random instances in the $Z_7$ ($Z_{150}$) case
    and $10$ independent runs per instance; error bars show one standard deviation. Insets in panels (a) and (b) show the measured runtime as a function of the number of steps.
    Panel (c) shows the difference $\Delta g$ between the optimality gaps obtained by SBM and SBQA as a function of instance size, with negative values favoring SBQA.
    The advantage of SBQA becomes visible beyond the $Z_{12}$ graph and grows as the problems become larger and sparser.
    }
    \label{fig:zephyr1}
\end{figure}

One of the major challenges for quantum hardware is the structure of couplings between individual qubits (physical or logical),
which defines the so-called working graph of the device. Restricted connectivity leads to the necessity of embedding~\cite{membedding}, which in turn
significantly degrades performance and limits the size of instances that can be studied. It is thus highly unlikely that near-term quantum annealers will
be able to demonstrate quantum advantage on problem classes that are not natively supported by their working graph. Because of that, the graphs
of D-Wave's Advantage and Advantage2 quantum annealers, namely the Pegasus and Zephyr architectures, have become a popular choice for assessing
the performance of Ising solvers.

In this section, we focus on the Zephyr graphs across four orders of magnitude in size, from $Z_1$ with
$N=47$ spins, through $Z_{12}$ with $N=4800$ spins (current generation Advantage2 QPU) up to $Z_{150}$ with $N=722400$ spins. Assuming
a doubling period of 2 years, a $Z_{150}$ working graph is expected to be within reach of quantum annealers in approximately a decade~\cite{veloxq2025}. For each 
of the graphs, we generate $20$ instances in the small regime ($Z_1 - Z_{20}$) and $5$ in the large
regime ($Z_{30}-Z_{150}$), with random couplings $J_{ij}$ and fields $h_i$ obtained from a QUBO matrix $Q_{ij}$ with entries drawn from a uniform
distribution over $[-1,1]$ (see Ref.~\cite{veloxq2025} for details). The reference energies are obtained by running large scale Simulated Annealing (SA) 
calculations, operating on timescales typically at least an order of magnitude larger than those required by SBM and SBQA.

\begin{figure}[t]
    \centering
    \includegraphics[width=\linewidth]{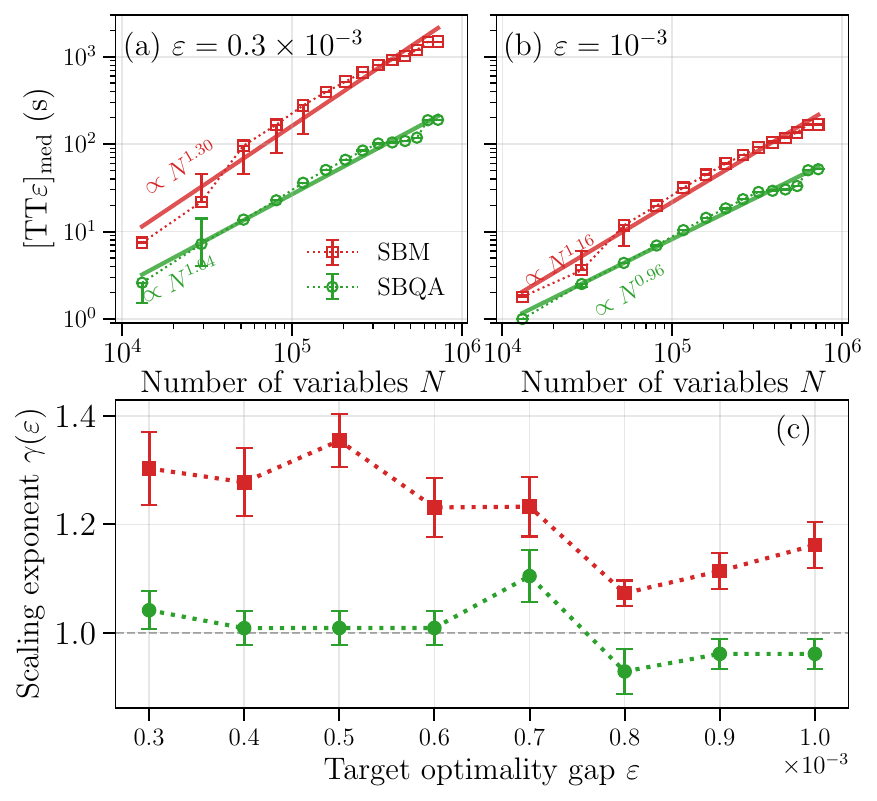}
    \caption{Scaling of time-to-epsilon with problem size for (a) $\varepsilon=0.3\times 10^{-3}$ and (b) $\varepsilon=10^{-3}$, in the range between
    $Z_{20}$ and $Z_{150}$ graphs. Data points show a linear trend in the log-log scale, indicating the expected power-law scaling, with the exponent
    $\gamma(\varepsilon)$ extracted from the fit.
    Panel (c) shows the dependence of the scaling exponent $\gamma$ on the target optimality gap $\varepsilon$, with SBQA consistently exhibiting
    a smaller exponent, and thus better scaling than SBM.
    }
    \label{fig:zephyr2}
\end{figure}

The results are shown in Figs.~\ref{fig:zephyr1} and~\ref{fig:zephyr2}. Figure~\ref{fig:zephyr1} shows that the advantage of SBQA over SBM emerges
in the regime of very large and very sparse problems, precisely where SBM is known to struggle. The solution quality, measured by the optimality gap relative
to the reference energy, is visibly improved already for instances with $N\sim 10^4$ variables and continues to improve with increasing size and decreasing density.
At the same time, the runtime overhead of SBQA becomes negligibly small as the number of variables grows.

In the moderate-to-large regime ($Z_{20}-Z_{150}$), we then study the scaling of time-to-epsilon with problem size. Here too SBQA exhibits better scaling than SBM,
as reflected in the dependence of the scaling exponent $\gamma$ on the target optimality gap $\varepsilon$ shown in panel (c) of Fig.~\ref{fig:zephyr2}. 

\subsubsection{Large-scale Quantum Annealing Correction (QAC) problems}

It is likely that genuine quantum advantage on near-term quantum annealers, if possible, will require some form of error correction.
A promising approach, called Quantum Annealing Correction (QAC), was put forth in Ref.~\cite{Pudenz2014}, and recently used
in an attempt to demonstrate quantum scaling advantage in Ref.~\cite{Lidar2025}. While it is clear that the QAC approach significantly
improves upon the performance of uncorrected quantum annealing, the question of whether it can actually achieve supremacy remains open. 
In particular, it was shown in Ref.~\cite{Pawlowski2025} that SBM surpasses the scaling of QAC on instances native to the topology
of the logical graph, even under a runtime definition designed to favor the annealer.

Here, we go beyond the sizes studied in Refs.~\cite{Lidar2025,Pawlowski2025} and consider logical QAC graphs of sizes from $L=20$ to $L=80$,
and random couplings drawn from the Sidon28 set $J_{ij} \in \{\pm 8/28, \pm 13/28, \pm 19/28, \pm 1\}$. The number of variables for these logical graphs
varies from $N=2380$ for $L=20$ to $N=38320$ for $L=80$. To run these instances on a real QPU, it would have to have approximately $N=1.5\cdot 10^5$ physical
qubits, arranged in the topology of Pegasus $P_{80}$ graph.

\begin{figure}[t]
    \centering
    \includegraphics[width=\linewidth]{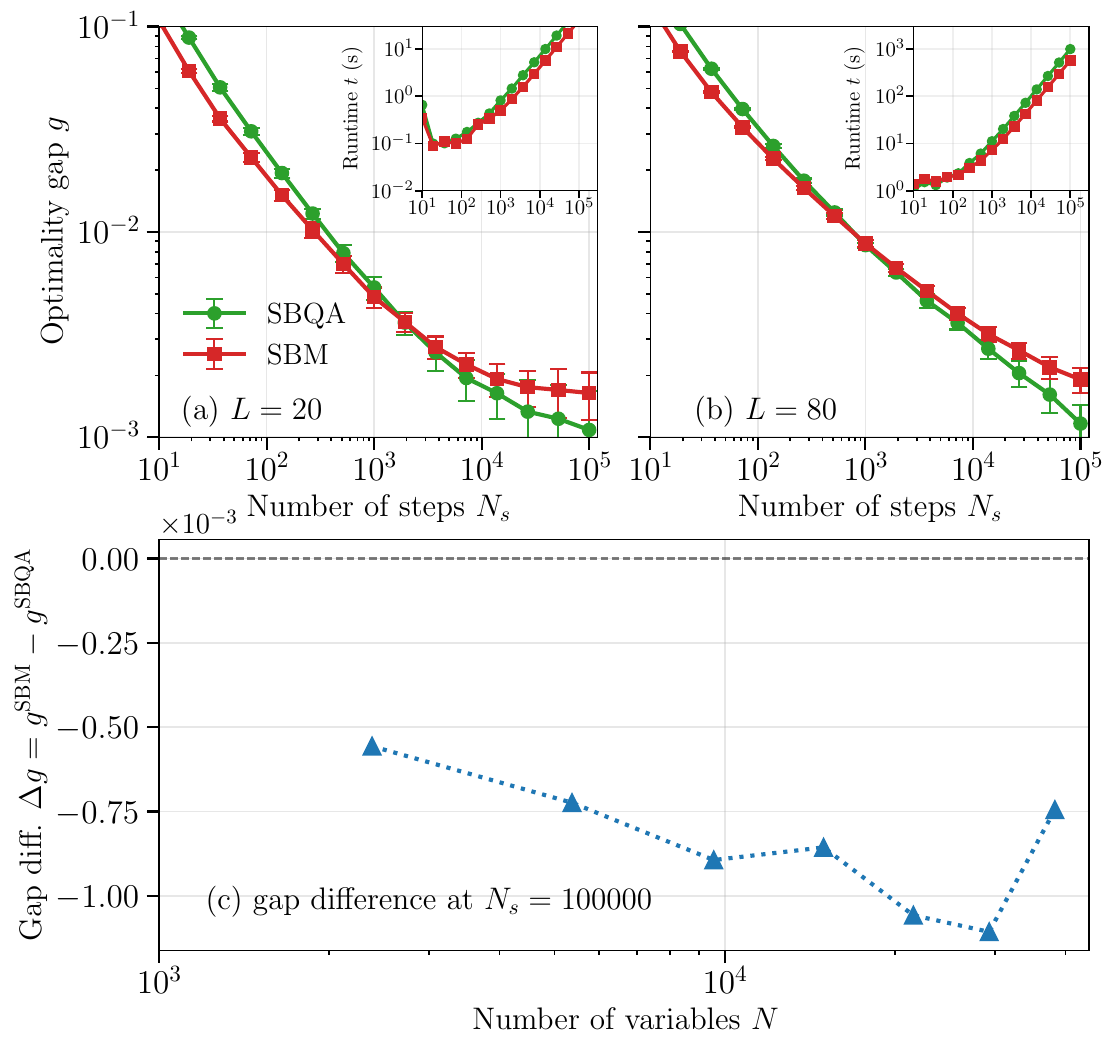}
    \caption{Average optimality gap $g$ as a function of the number of steps for Sidon28 instances defined on the QAC logical graph with size parameter
    (a) $L=20$ and (b) $L=80$. Results are averaged over an ensemble of $10$ random instances for each size and $10$ independent runs per instance;
    error bars show one standard deviation. Insets in panels (a)~and (b) show the measured runtime as a function of the number of steps.
    Panel (c) shows the difference $\Delta g$ between the optimality gaps obtained by SBM and SBQA at fixed $N_s=10^5$, with negative values favoring SBQA.
    For a sufficiently large number of steps, the advantage of SBQA becomes significant across all studied sizes.
    }
    \label{fig:large_lidar_1}
\end{figure}

\begin{figure}[t]
    \centering
    \includegraphics[width=\linewidth]{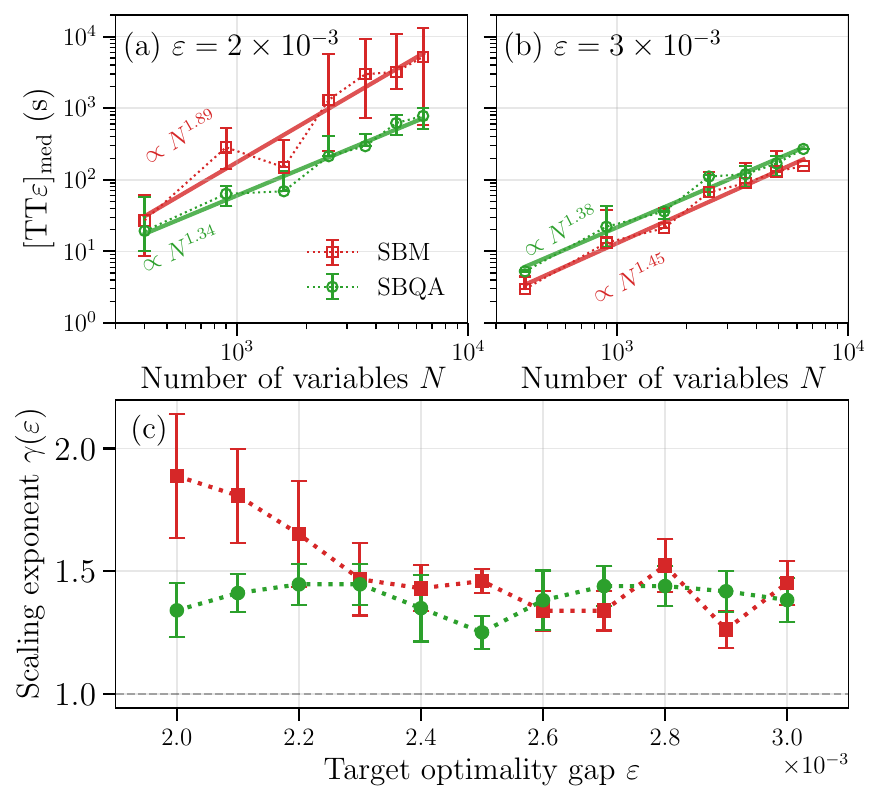}
    \caption{Scaling of time-to-epsilon with problem size for (a) $\varepsilon=2\times 10^{-3}$ and (b) $\varepsilon=3 \times 10^{-3}$, in the range between
    $L=20$ and $L=80$ QAC instances. Bottom panel (c) shows the exponent $\gamma(\varepsilon)$ extracted from the power-law fit. 
    Both solvers perform comparably for the optimality gap range \mbox{$0.23\% \lesssim \varepsilon \lesssim 0.3\%$}, but for stricter target gaps
    SBQA starts to show a pronounced advantage. 
    }
    \label{fig:large_lidar_2}
\end{figure}

The results are shown in Figs.~\ref{fig:large_lidar_1} and~\ref{fig:large_lidar_2}. We find that for a small number of steps SBM outperforms SBQA, but after $N_s \simeq 10^3$, 
the advantage of SBQA emerges and grows with increasing system size. This translates to a~visible improvement in the scaling of time-to-epsilon for
the strictest achievable target optimality gaps $\varepsilon \simeq 0.2\%$. We thus conclude that SBQA is capable of raising the bar further for quantum advantage with QAC,
and it will certainly be interesting to see how it compares to the performance of future quantum annealers with QAC.

\subsubsection{2D and 3D tile planting}
\begin{figure}[htbp]
    \centering
    \includegraphics[width=\linewidth]{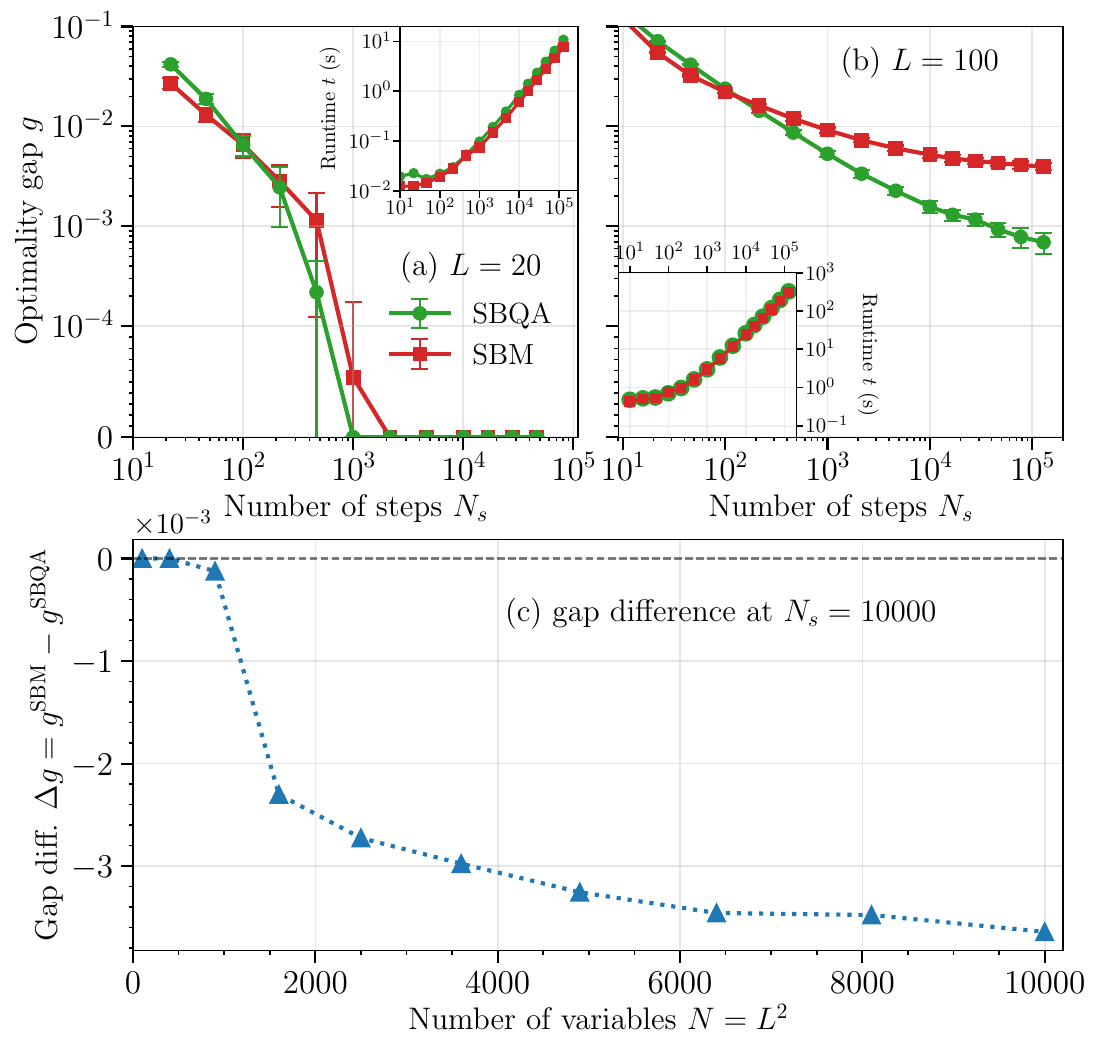}
    \caption{Average optimality gap $g$ as a function of the number of steps for instances on a square lattice with linear dimension
    (a) $L=20$ and (b) $L=100$. The results are averaged over an ensemble of $10$ random instances for each size,
    and $10$ independent runs per instance. Error bars correspond to one standard deviation. Insets in panels (a) and (b) show the 
    measured runtime as a function of the number of steps. Panel (c) shows the difference $\Delta g$ between the optimality gaps obtained by SBM and SBQA,
    as a function of instance size, negative values indicating better performance of SBQA. For the smallest instance sizes $L\leq 30$ the performance of SBM and SBQA
    is comparable, with both of them producing the planted ground state, or a very close solution, in the majority of runs. 
    For larger sizes, however, a significant gap opens between the two algorithms, with SBQA consistently outperforming SBM, and the advantage is growing
    with increasing problem size.
    }
    \label{fig:tile_planting_2D_1}
\end{figure}

\begin{figure}[t]
    \centering
    \includegraphics[width=\linewidth]{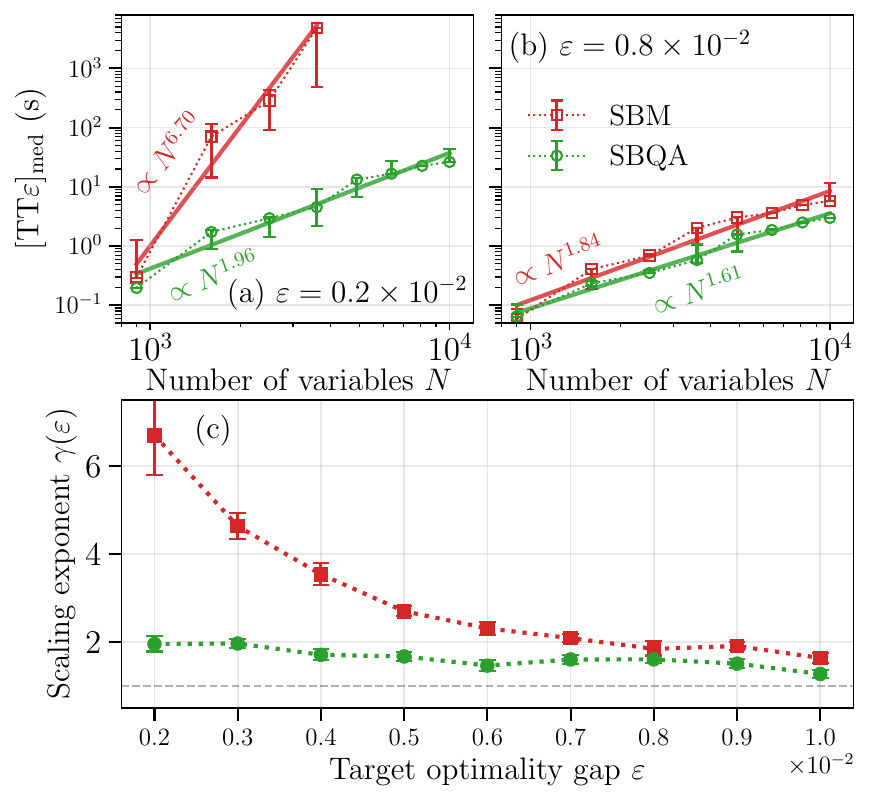}
    \caption{Scaling of time-to-epsilon with problem size for (a) $\varepsilon=0.2\%$ and (b) $\varepsilon=0.8\%$, in the range between
    $L=30$ and $L=100$ square lattices. Bottom panel (c) shows the exponent $\gamma(\varepsilon)$ extracted from the power-law fit. 
    In this case, the advantage of SBQA over SBM is even more pronounced than in the case of Zephyr, with the SBM scaling exponent seemingly
    already starting to diverge towards the exponential regime of $\mathrm{TTS}$, while SBQA remains firmly in the power-law regime.
    This result can be attributed to the extremely sparse nature of square lattice problems, and further supports the conclusion that 
    inter-replica coupling improves the performance in such regimes.
    }
    \label{fig:tile_planting_2D_2}
\end{figure}

In the context of $\mathrm{TT}\varepsilon$ studies, it is desirable to have access to instances spanning a wide range of sizes
and hardness, with known ground state energies, to avoid the pitfalls of extrapolating from small sizes and relying on suboptimal reference energies.
We thus turn our attention to the so-called planted solution instances, which are constructed in a way that guarantees that the ground state energy is known,
regardless of the problem size.

More precisely, we consider tile-planting instances defined on square (2D)~\cite{Perera2020} and
cubic (3D)~\cite{Hamze2018} lattices. They are constructed by tiling the lattice with randomly chosen plaquettes drawn from a predefined
set of types with varying degrees of frustration, which allows the hardness of the problem to be tuned through the tile composition.
This feature makes them a particularly useful benchmarking ground for heuristic optimization algorithms. In fact, it has already been shown that SBM tends to underperform
on these instances~\cite{veloxq2025}, in particular in comparison to other approaches based on non-linear dynamical systems~\cite{Katzgraber2025}.

In the 2D case, we consider instances of \(C_2-C_4\) type, with probability of \(C_2\) tile being \(p_2 = 0.8\), which pushes them into the hard regime~\cite{Perera2020}. 
For the 3D case, we use the \texttt{gallus46} tile set, with parameter $p_6=0.6$~\cite{Hamze2018,Gangat2026}.

\begin{figure}[htbp]
    \centering
    \includegraphics[width=\linewidth]{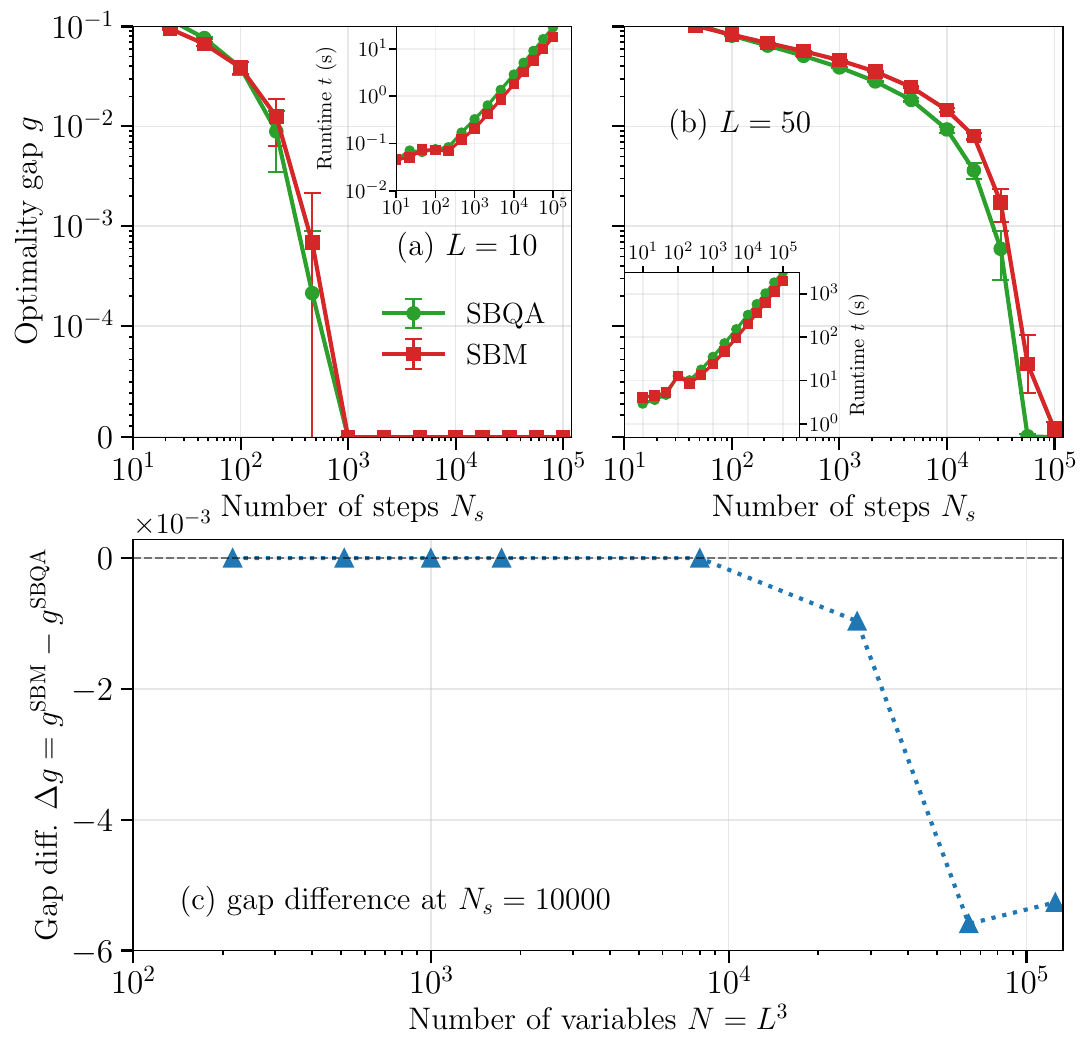}
    \caption{Average optimality gap $g$ as a function of the number of steps for instances on a cubic lattice with linear dimension
    (a) $L=10$ and (b) $L=50$. Results are averaged over an ensemble of $10$ random instances for each size and $5$ independent runs per instance;
    error bars show one standard deviation. Insets in panels (a) and (b) show the measured runtime as a function of the number of steps.
    Panel (c) shows the difference $\Delta g$ between the optimality gaps obtained by SBM and SBQA as a function of instance size, with negative values favoring SBQA.
    The advantage of SBQA emerges for $L>20$, but it is less pronounced than in the square-lattice case.
    }
    \label{fig:tile_planting_3D_1}
\end{figure}

\begin{figure}[t]
    \centering
    \includegraphics[width=0.7\linewidth]{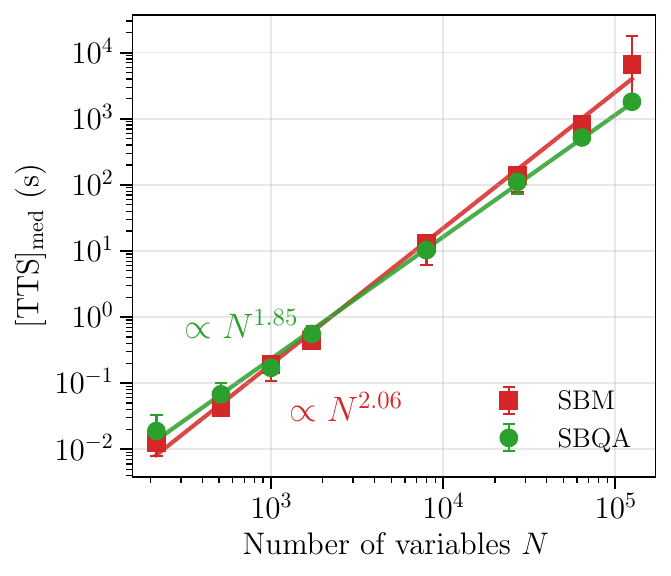}
    \caption{
        Scaling of time-to-solution with problem size in the range between
    $L=10$ and $L=50$ for cubic-lattice tile planting. While $[\mathrm{TTS}]_{\rm med} \sim \mathrm{exp}(N)$ scaling is typically expected,
    we observe polynomial scaling for both SBM and SBQA, with SBQA yielding the smaller exponent.
    As already noticed in Fig.~\ref{fig:tile_planting_3D_1}, the difference between SBM and SBQA shrinks considerably in the case of 3D tile planting,
    but is still enough to see a measurable improvement.
    }
    \label{fig:tile_planting_3D_2}
\end{figure}

We show the results for 2D tile planting in Figs.~\ref{fig:tile_planting_2D_1}-\ref{fig:tile_planting_2D_2}, and for 3D tile planting in Figs.~\ref{fig:tile_planting_3D_1}-\ref{fig:tile_planting_3D_2}. 

In the 2D case, replica coupling in SBQA leads to a significant improvement over SBM, which is especially pronounced for larger instance sizes. Since these
instances are on the extremely sparse end of the density spectrum, this constitutes strong evidence for our claim that SBQA remedies one of the known weaknesses of SBM, 
namely poor performance on very sparse problems. 

The effect is visible both at fixed computational cost (number of steps $N_s$) and in the time-to-epsilon scaling. In particular, in Fig.~\ref{fig:tile_planting_2D_2} SBQA yields systematically
smaller \(\mathrm{TT}\varepsilon\) and a lower effective scaling exponent \(\alpha(\varepsilon)\) across the considered \(\varepsilon\) range, while SBM moves toward substantially
worse scaling for stricter targets. This indicates that inter-replica coupling improves not only individual values of $\mathrm{TT}\varepsilon$ (fixed quality solution is reached faster),
but also the size scaling in the hard 2D regime.

For 3D tile planting, the qualitative picture is different, and both algorithms perform comparably well. In particular, even for the largest studied size of $L=50$
both SBM and SBQA yield a nonzero probability of finding the planted ground state. This allows us to focus only on the $\varepsilon=0.0\%$ limit, i.e. the
original time-to-solution metric $\mathrm{TTS}$~\cite{Troyer2014}.

In Fig.~\ref{fig:tile_planting_3D_2}, we show that, contrary to the typically expected exponential scaling,
both solvers exhibit polynomial scaling across the presently accessible size range, with SBQA yielding a slightly smaller exponent $\gamma$ than SBM.
This behavior may still be a finite-size effect, and the expected exponential scaling may emerge for larger instances, but the observed regime is sufficient for comparing the two solvers.

It is also illuminating to compare this result with the recent findings of Ref.~\cite{Gangat2026}, where the same class of instances was found to be among the hardest
for the linear-time heuristic proposed therein, with the estimated $\varepsilon_{\mathrm{lin}}\simeq 7.5\%$. This positions the 3D tile planting instances
as an interesting test-bed for optimization heuristics.

Overall, the tile-planting benchmarks support the central claim of this work: SBQA provides the most significant gains exactly where SBM is weakest, i.e. on large and sparse problems.
While this gain diminishes for denser problems, it can still be observed for a sufficiently large number of variables, and is not accompanied by any significant runtime overhead. 

\subsection{Instances on current-generation quantum hardware \label{sec:hardware}}
While the asymptotic scaling of time-to-epsilon is an important theoretical benchmark, and, when done properly, can disentangle the algorithmic properties from implementation details, 
it is not always the case that an algorithm with better asymptotic scaling will be better for finite-size problems of practical relevance. 
In particular, near-term quantum devices operate in a regime where problem sizes, connectivity constraints, and hardware-specific overheads play an important role in practical performance.
Therefore, it is desirable to complement asymptotic analyses with benchmarks on instances that are directly compatible with the architectures of current-generation quantum hardware.

In this section, we shift our focus to such practically relevant regimes, considering problem classes that can be natively solved on existing quantum platforms.
This includes 3D Ising spin-glass instances defined on the Pegasus topology of the D-Wave quantum annealer~\cite{King2023,Chowdhury2025}, as well as higher-order binary optimization
problems derived from the heavy-hex topology of IBM processors. These benchmarks allow us to assess SBQA not only as an algorithm with favorable asymptotic scaling,
but also as a~competitive heuristic under realistic conditions.

\subsubsection{3D spin glasses on D-Wave quantum annealer}
We start with benchmarks on 3D spin glasses, with couplings distributed according to the standard normal distribution.
This is a modification of the instances already studied in the context of combinatorial optimization and quantum
annealing, for which both the D-Wave annealer and DTSQA performed very well~\cite{King2023,Chowdhury2025}. The logical topology of the
instances is a cubic lattice with periodic boundaries in two directions and an open boundary in the third direction.
Such graphs have a simple embedding into the Pegasus architecture, with the $P_M$ graph accommodating
a lattice of dimensions at most \((M-1)\times (M-1) \times 12\), using chains of fixed length \(2\)~\cite{Boothby2020}.
We consider cubic lattices of size \(L^3\) for \(L=6, 8, 10, 12\) and \(L^2 \times 12\) for \(L=15\). For the reference energies,
we used the DTSQA algorithm run with $5 \times 10^5$ steps. We note that DTSQA requires graph coloring before the computation
can be run, and this preprocessing can be performed once for a given instance topology. In our implementation, we use the DSATUR algorithm and exclude the coloring time from the runtime measurements.

\begin{figure}[t]
		    \centering
		    \includegraphics[width=\linewidth]{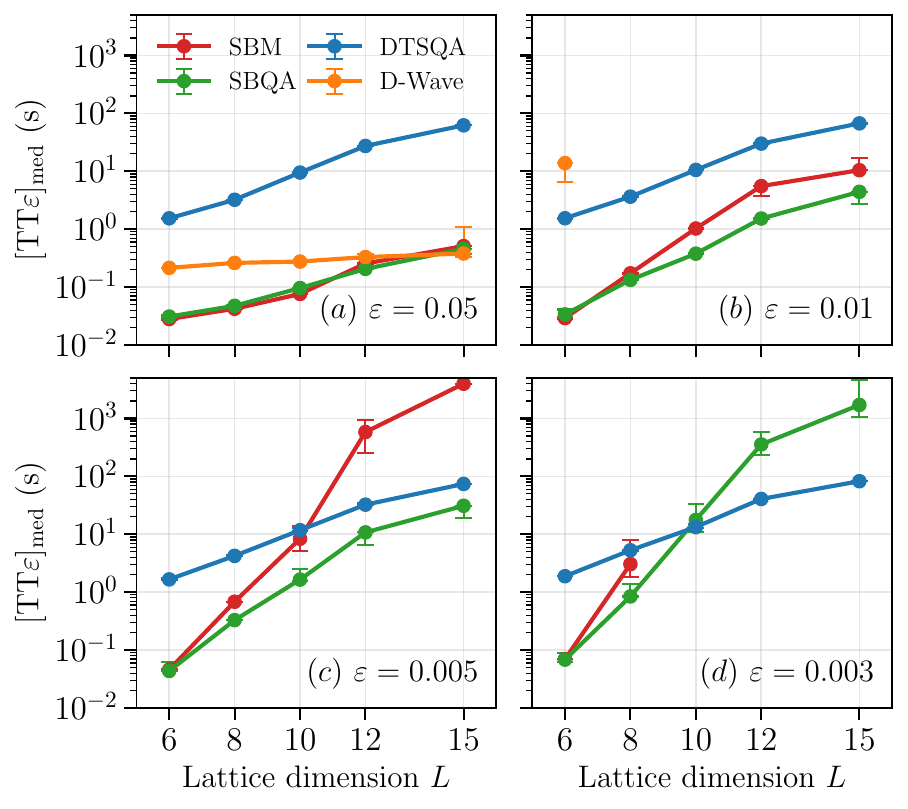}
	    \caption{Time-to-epsilon benchmark for 3D spin glass instances embedded into the Pegasus \(P_{16}\) graph,
	        with sizes ranging from \(L=6\) with \(2\times 6^3=432\) variables to \(L=15\) with \(15^2\times 12 \times 2=5400\).
	        The target optimality gaps are set to (a) \(0.05\), (b) \(0.01\), (c) \(0.005\), and (d) \(0.003\).
	        SBQA matches or outperforms SBM across the shown range and achieves smaller \([\mathrm{TT}\varepsilon]_{\rm med}\) values than DTSQA except for the largest instances at the strictest thresholds.
	        D-Wave performs reasonably well only for the easiest target and quickly falls behind as the target is made stricter.}
    \label{fig:3D_spin_glass_embedded}
\end{figure}

In Fig.~\ref{fig:3D_spin_glass_embedded}, we compare the SBQA time-to-epsilon values for embedded instances of different sizes against our baseline SBM,
DTSQA, and the D-Wave Advantage 4.1 quantum annealer. For each data point, we consider \(50\) random instances (\(10\) for D-Wave) and \(5\) independent
runs (\(10\) for D-Wave) per instance to estimate the success probability and average runtime (QPU access time for the annealer). As usual, the time-to-epsilon values are
minimized over the number of steps of the algorithm, or the annealing time in the case of D-Wave.

Despite previous results showing strong performance of quantum annealing on this topology, changing the coupling distribution severely degrades its performance. As a result,
\([\mathrm{TT}\varepsilon]_{\rm med}\) remains finite only for the largest optimality gap considered here, \(\varepsilon=0.05\). By contrast, SBQA outperforms SBM for all
optimization targets except the easiest one, where the two algorithms perform identically. This reflects the fact that replica interaction improves solution quality at a fixed
number of steps while introducing very little runtime overhead.

SBQA also outperforms DTSQA in most cases. The main exception is the regime of the largest instances and the strictest optimization targets, where only DTSQA is able to
consistently reach the target and avoid the penalty associated with very small success probabilities. It does so, however, at the cost of significantly longer runtimes.
Additional plots decomposing time-to-epsilon into solution-quality and runtime components are provided in Sec.~\ref{sec:3d_supplemental} of Ref.~\cite{supmat}.

\begin{figure}[htbp]
	    \centering
	    \includegraphics[width=\linewidth]{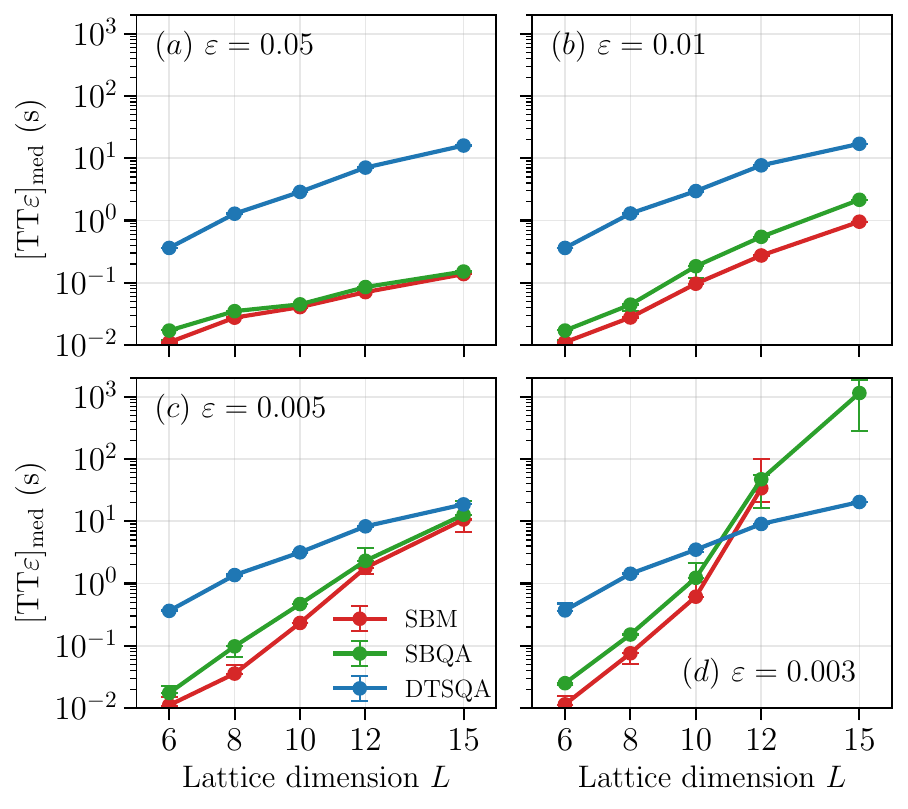}
	    \caption{Time-to-epsilon benchmark for logical 3D spin glass instances,
	        with sizes ranging from \(L=6\) with \(6^3=216\) variables to \(L=15\) with \(15^2\times 12 =2700\).
	        The target optimality gaps are set to (a) \(0.05\), (b) \(0.01\), (c) \(0.005\), and (d) \(0.003\).
	        SBQA performs very similarly to SBM across all targets, with a small runtime penalty due to inter-replica coupling.
	        Compared with DTSQA, SBQA again achieves smaller \([\mathrm{TT}\varepsilon]_{\rm med}\) values except for the largest instances at the strictest thresholds.}
    \label{fig:3D_spin_glass_logical}
\end{figure}

The results for the logical instances are shown in Fig.~\ref{fig:3D_spin_glass_logical}. Here, both SBQA and SBM yield solutions of very similar quality (especially for larger instances),
so the SBM retains a slight edge in time-to-epsilon due to the lower computational cost per step. 
Similarly to the embedded instances, DTSQA falls behind due to its longer runtimes, except for the largest instances and strictest optimization targets.

\subsubsection{Higher-order binary optimization on heavy-hex topology}

\begin{figure}[htbp]
    \centering
    \includegraphics[width=\linewidth]{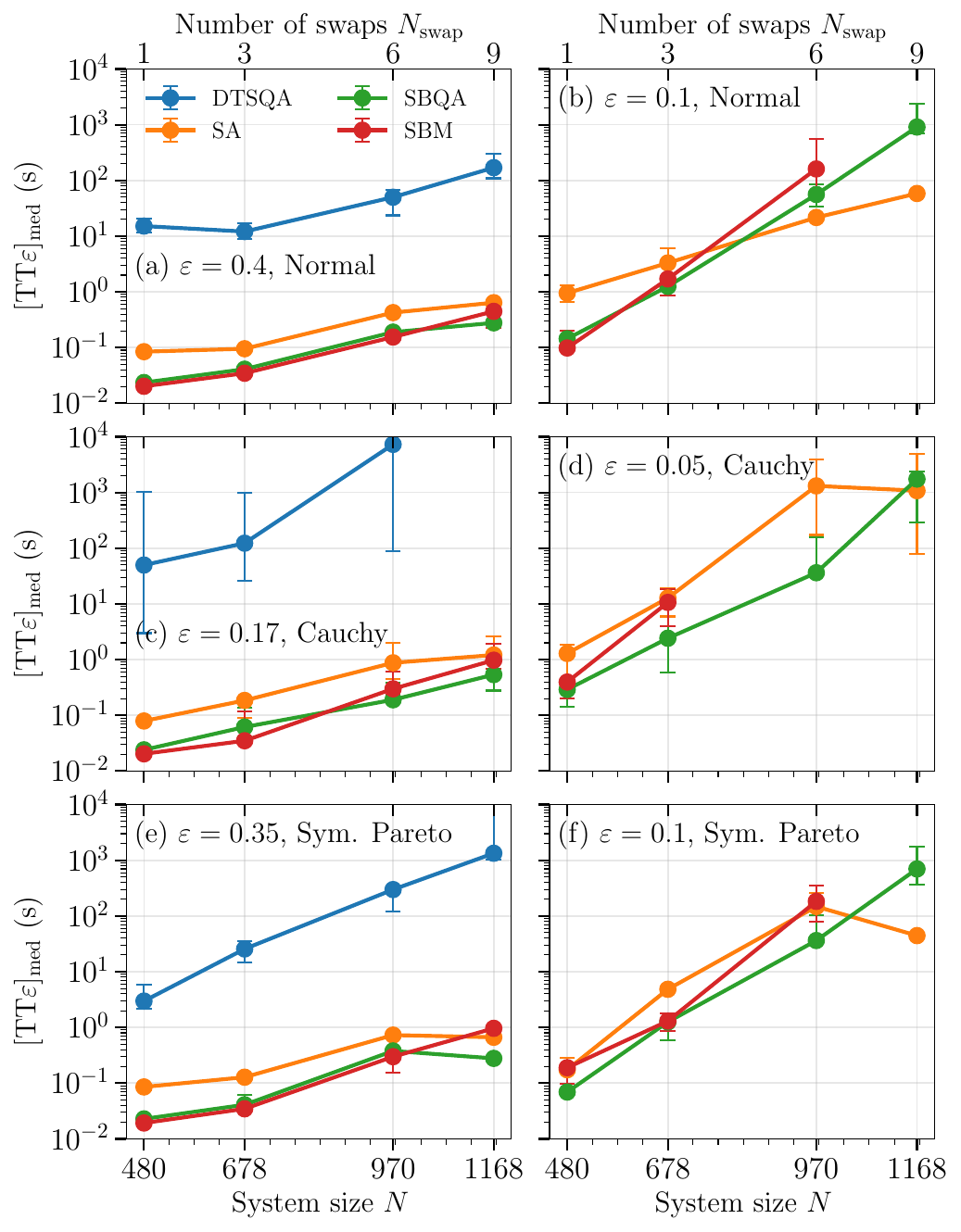}
    \caption{Time-to-epsilon benchmark
        for HUBO instances constructed from the heavy-hexagon topology of IBM quantum processors,
        with \(N_{\rm swap} \in \{1, 3, 6, 9\}\) and number of variables ranging from \(N = 480\) to \(N = 1168\) after
        reduction to QUBO form. Rows of panels correspond to different coupling distributions: (a-b) Normal, (c-d) Cauchy,
        and (e-f) Symmetrized Pareto, whereas columns correspond to different optimality targets, with first column containing
        ``easy'' targets and second column to ``hard'' targets (dependent on the coupling distribution).
        For each data point, \(20\) random instances and \(10\) independent runs per instance were used to estimate the $[\mathrm{TT}\varepsilon]_{\rm med}$ values,
        which were minimized over the number of steps of the algorithm.
DTSQA performs poorly, struggling to leverage the graph structure for such complex topologies, and is only able to reach the ``easy'' optimization targets,
        completely failing for the stricter ones in the second column. On the other hand, SBQA continues to perform well, matching SBM and slightly outperforming
        SA on ``easy'' targets. For ``hard'' targets, SBQA significantly outperforms SBM for larger problem sizes ($N_{\rm swap} = 6,9$), and loses to SA
        only for the two largest instances with Normal couplings (panel (b)) and the largest instance with Pareto couplings (panel (f)).
        }
    \label{fig:hubo}
\end{figure}
For our final set of benchmarks, we consider a different class of problems, namely higher-order binary optimization (HUBO) problems constructed from
the heavy-hexagon topology of IBM quantum processors~\cite{heavyhex}. These problems have recently emerged as a challenging benchmark set for optimization routines,
and have been used in claims of runtime quantum advantage for gate-based quantum computing~\cite{Chandarana2025}, claims that have been subsequently challenged
by highly optimized classical heuristics~\cite{Tuziemski2025}. The Hamiltonian of the problem in the Ising spin variables reads:

\begin{equation}
    H = \sum_{(m,n)\in G_2} J_{mn} s_m s_n + \sum_{(l,m,n)\in G_3} K_{lmn} s_l s_m s_n,
    \label{eq:hubo}
\end{equation}
where \(G_2\) and \(G_3\) are the sets of 2-body and 3-body interactions.
The starting point of the construction is \mbox{a heavy-hexagon} graph of size \(N=156\), corresponding to
the topology of IBM Heron QPUs, from which sets of independent 2-body and 3-body interactions (which can be executed simultaneously on a real QPU)
are constructed via graph coloring. A fixed number of such sets, $S_{2q}$ and $S_{3q}$, are then included into \(G_2\) and \(G_3\), respectively.
Finally, using one of the two-body sets, a SWAP operation is carried out between connected pairs of qubits, which modifies the underlying graph. This procedure
can be repeated \(N_{\rm swap}\) times, increasing the number of interactions with each iteration. With precomputed topology,
an ensemble of random HUBO instances is generated by drawing the coupling strengths \(J_{mn}\) and \(K_{lmn}\) from suitable distributions.
See the Appendix in Ref.~\cite{Tuziemski2025} for technical details on the construction of these HUBO instances, and repository~\cite{github}
for the implementation used in this work. To obtain the reference energies, we used CPLEX optimizer running on native HUBO instances, as well as
a variant of higher-order formulation of SBM~\cite{Kanao2022}, both of which were able to produce considerably better solutions than the reduction-based approaches, 
especially for $N_{\rm swap} > 1$.

The value of $N_{\rm swap}$ directly controls the depth of the quantum circuit required to solve the problem on a gate-based quantum computer, and simultaneously
the hardness of the resulting HUBO instance for classical heuristics. In particular, for $N_{\rm swap} > 1$ the resulting circuit depth for the approach used
by the authors of Ref.~\cite{Chandarana2025} already exceeds the capabilities of IBM's QPU. This feature makes these instances particularly interesting for benchmarking purposes.
In our tests, we use $S_{2q}=1$, $S_{3q}=6$, $N_{\rm swap}\in \{1,3,6,9\}$ and couplings drawn from Cauchy, Normal and Symmetrized Pareto distributions.
Even though there exist specialized methods for solving HUBO problems directly~\cite{Kanao2022,romero2024}, here we focus on comparing SBQA against standard SBM,
SA and DTSQA, which require the reduction of the HUBO problem to QUBO form. To this end we use the D-Wave's \emph{dimod} package and a standard reduction method
based on introducing auxiliary variables implemented therein~\cite{dimod-make-quadratic, Rosenberg1975}. It uses a greedy heuristic to minimize the number of
auxiliary variables required for the reduction. See Table~\ref{tab:optimal_penalties_comparison} in Ref.~\cite{supmat} for additional details on the penalty coefficient.

The results of the time-to-epsilon benchmarks are shown in Fig.~\ref{fig:hubo}, where rows of panels correspond to different coupling distributions,
and columns to different optimality targets, with the first column containing ``easy'' targets and the second column ``hard'' targets.
The first clear observation is the relatively weak performance of DTSQA. It reaches the ``easy'' optimization targets only in part of the benchmark set and fails completely for the stricter targets in the second column.
Although DTSQA performed excellently in the 3D spin-glass experiments, the data for QUBO instances obtained from HUBO reduction suggest that it is not well suited to such complex topologies.

By contrast, both SBQA and SBM continue to perform well. SBQA essentially matches SBM on the ``easy'' targets and significantly outperforms it for the larger reduced-problem sizes ($N_{\rm swap}=6,9$) on the ``hard'' targets.
Simulated Annealing, our reference temperature-based method, falls behind on the ``easy'' targets because of its longer runtimes, but still produces very good solutions on the ``hard'' targets.
This leads to lower $[\mathrm{TT}\varepsilon]_{\rm med}$ values in two cases: the two largest instances with Normal couplings (panel (b)) and the largest instance with Pareto couplings (panel (f)).

Overall, these results show that adding inter-replica coupling in SBQA does not compromise versatility, and allows the method to remain effective across a wide range of problem types and topologies.

\section{Summary and outlook\label{sec:conclusions}}

In this work, we introduced \emph{Simulated Bifurcation Quantum Annealing} (SBQA) and demonstrated that incorporating controlled 
inter-replica interactions into the simulated bifurcation framework leads to measurable and consistent performance improvements on 
problem classes where standard SBM is known to struggle, particularly sparse and rugged energy landscapes. At the same time, 
this modification preserved its computational efficiency and parallelism.

Using a rigorous time-to-epsilon benchmarking methodology with real runtime
measurements, we showed that SBQA achieves a favorable balance between solution
quality and runtime across a diverse set of benchmarks, including large and
sparse problems as well as moderately sized instances relevant for
current-generation quantum hardware. Across the tested benchmarks, SBQA
systematically improves on standard SBM in the sparse and rugged regimes that
motivate the method, while remaining broadly effective across other instance
families considered in this work.

Beyond its immediate empirical performance, SBQA illustrates a broader
algorithmic principle: modest, carefully engineered modifications inspired by
quantum dynamics can yield practically significant gains in classical
heuristics. In regimes where quantum--classical performance comparisons hinge on
narrow margins, such improvements can materially strengthen the classical
baseline used for evaluation.

Several directions for future work remain open. On~the algorithmic side, further
exploration of adaptive or problem-structure-aware replica coupling schedules
may yield additional performance gains, as may extensions of SBQA to alternative
cost-function encodings or constrained optimization settings. From a
benchmarking perspective, applying SBQA to larger-scale industrial instances and
real-world optimization workloads would help clarify its practical impact beyond
synthetic benchmarks. Finally, SBQA provides a useful and physically motivated
classical baseline for future studies comparing quantum optimization devices
with classical heuristics on sparse and rugged benchmark families.

\section*{Acknowledgments}
This project was supported by the National Science Center (NCN), Poland, under
Projects: Sonata Bis~10, No.~2020/38/E/ST3/00269 (B.G.) and Sonata Bis~15, No.~2025/58/E/ST6/00422 ({\L}.P.). Quantumz.io Sp. \mbox{z o.o.}
acknowledges support received from The National Centre for Research and
Development (NCBR), Poland, under Project No.~POIR.01.01.01-00-0061/22.

\bibliography{lit}

\newpage
\phantom{a}
\newpage

\setcounter{figure}{0}
\setcounter{equation}{0}
\setcounter{page}{1}
\setcounter{section}{0}

\renewcommand{\thetable}{S\arabic{table}}
\renewcommand{\thefigure}{S\arabic{figure}}
\renewcommand{\thepage}{S\arabic{page}}
\renewcommand{\thesection}{S\arabic{section}}
\renewcommand{\theequation}{S\arabic{equation}}

\onecolumngrid

\begin{center}
	{\large \bf Supplemental Material:\\
		Simulated Bifurcation Quantum Annealing}
	\vspace{0.3cm}
\end{center}

In the Supplemental Material we present a concise derivation of the Simulated Quantum Annealing Hamiltonian, as well as fine-grained
analysis of the benchmarks on hardware-compatible instances, including 3D spin glasses on the Pegasus topology and HUBO instances derived from the heavy-hex topology of IBM quantum processors.

\vspace{0.6cm}
\section{Derivation of the inter-replica coupling in Simulated Quantum Annealing \label{sec:sqa_derivation}}
Simulated Quantum Annealing requires mapping a $d$-dimensional quantum system onto a $(d+1)$-dimensional classical system~\cite{Suzuki1976PIMC}. We consider the quantum Hamiltonian:
\begin{equation}
	H_{\rm Q} = -\sum_{i<j} J_{ij} \sigma^{z}_{i}\sigma^{z}_{j} - \Gamma_{x}\sum_{i}\sigma^{x}_{i}
\end{equation}
with partition function $Z_{\rm Q} = \Tr[e^{-\beta H_{\rm Q}}]$. Due to non-commutativity between $H_z = -\sum_{i<j} J_{ij} \sigma^{z}_{i}\sigma^{z}_{j}$ 
and $H_x = -\Gamma_{x}\sum_{i}\sigma^{x}_{i}$, we employ the Suzuki-Trotter decomposition to construct the equivalent classical representation:
\begin{equation}
	e^{-\beta (H_z + H_x)} = \lim_{R\to\infty} \left[ e^{-\frac{\beta}{R} H_z} e^{-\frac{\beta}{R} H_x} \right]^R,
\end{equation}
where $R$ represents the number of imaginary-time replicas. This decomposition introduces $R$ copies of the system along an extra dimension. 
Inserting complete sets of $\sigma^z$-basis states $\{\ket{\boldsymbol{\sigma}^k}\}$ with periodic boundary conditions 
$\boldsymbol{\sigma}^{R+1} = \boldsymbol{\sigma}^1$, we obtain:
\begin{equation}
	Z_{\rm Q} = \lim_{R\to\infty} \Tr \left( \prod_{k=1}^R e^{-\frac{\beta}{R} H_z} e^{-\frac{\beta}{R} H_x} \right)
	= \lim_{R\to\infty} \sum_{\{\boldsymbol{\sigma}^k\}} \prod_{k=1}^R \bra{\boldsymbol{\sigma}^k}  e^{-\frac{\beta}{R} H_z}
    e^{-\frac{\beta}{R} H_x} \ket{\boldsymbol{\sigma}^{k+1}}.
\end{equation}
Since $H_z$ is diagonal in the $\sigma^z$-basis:
\begin{equation}
	\bra{\boldsymbol{\sigma}^k} e^{-\frac{\beta}{R} H_z} \ket{\boldsymbol{\sigma}^{k+1}} =
    \exp\left( \frac{\beta}{R} \sum_{i<j} J_{ij} \sigma_i^k \sigma_j^k \right) \delta_{\boldsymbol{\sigma}^k, \boldsymbol{\sigma}^{k+1}},
\end{equation}
whereas the non-diagonal term factors by site:
\begin{equation}
	\bra{\boldsymbol{\sigma}^k} e^{-\frac{\beta}{R} H_x} \ket{\boldsymbol{\sigma}^{k+1}} = 
    \prod_i \bra{\sigma_i^k} \exp\left( \frac{\beta \Gamma_x}{R} \sigma_i^x \right) \ket{\sigma_i^{k+1}}.
\end{equation}
The single-site matrix element evaluates to:
\begin{align}
	\bra{\sigma_i^k} e^{\theta \sigma^x} \ket{\sigma_i^{k+1}} & =  \cosh(\theta) \delta_{\sigma_i^k, \sigma_i^{k+1}} + \sinh(\theta) \delta_{\sigma_i^k, -\sigma_i^{k+1}} \nonumber \\
	                                                          & = \exp(a + b \sigma_i^k \sigma_i^{k+1}),
\end{align}
where $\theta = \beta \Gamma_x / R$, and parameters $a$ and $b$ are defined as:
\begin{equation}
	\begin{cases}
		\exp(a+b) = \cosh(\theta) \\
		\exp(a-b) = \sinh(\theta)
	\end{cases}
\end{equation}
which can be solved to yield:
\begin{align}
	a & = \frac{1}{2} \ln \left( \cosh(\theta) \sinh(\theta) \right) = \frac{1}{2} \ln \left(\frac{1}{2}\sinh(2\theta)\right) \\
	b & = \frac{1}{2} \ln \left( \frac{\cosh(\theta)}{\sinh(\theta)} \right) = \frac{1}{2} \ln \left( \coth(\theta) \right).
\end{align}
The spin-independent prefactor can be neglected if we are only interested in the emergent classical Hamiltonian.
Combining both terms yields:
\begin{equation}
	Z_{\rm Q} = \lim_{R\to\infty} \sum_{\{\sigma_{i,k}\}} \exp\Bigg[ \frac{\beta}{R} \sum_{k=1}^R \sum_{i<j} J_{ij} \sigma_{i}^k\sigma_j^k + 
    \underbrace{\frac{1}{2} \ln \coth \left( \frac{\beta \Gamma_x}{R} \right)}_{b} \sum_{k=1}^R \sum_i \sigma_i^k\sigma_i^{k+1} \Bigg]
\end{equation}
Identifying the exponent as $-\beta H_{\rm C}$ gives:
\begin{equation}
	H_{\rm C}  = -\frac{1}{R} \sum_{k=1}^R \sum_{i<j} J_{ij} \sigma_{i}^{k}\sigma_{j}^{k} - \frac{K}{\beta} \sum_{k=1}^R \sum_i \sigma_{i}^{k}\sigma_{i}^{k+1}
	\label{eq:dt-sqa-hamiltonian}
\end{equation}
with inter-replica coupling:
\begin{equation}
	J_\perp \equiv \frac{K}{\beta} = \frac{1}{2\beta} \ln \coth \left( \frac{\beta \Gamma_x}{R} \right) = -\frac{1}{2\beta} \ln \tanh \left( \frac{\beta \Gamma_x}{R} \right)
	\label{eq:dt-sqa-coupling}
\end{equation}
This is the exact SQA-derived coupling for a given transverse field $\Gamma_x$.
In the time-dependent SBQA implementation discussed in the main text, we use the
regularized schedule
$\Gamma_x(t)=\Gamma_x(0)\left[(1-t/T)^{\alpha}+10^{-5}\right]$ to keep the
endpoint finite.

\section{Additional results for 3D spin glass benchmark\label{sec:3d_supplemental}}
\begin{figure}[htbp]
    \centering
    \includegraphics[width=\linewidth]{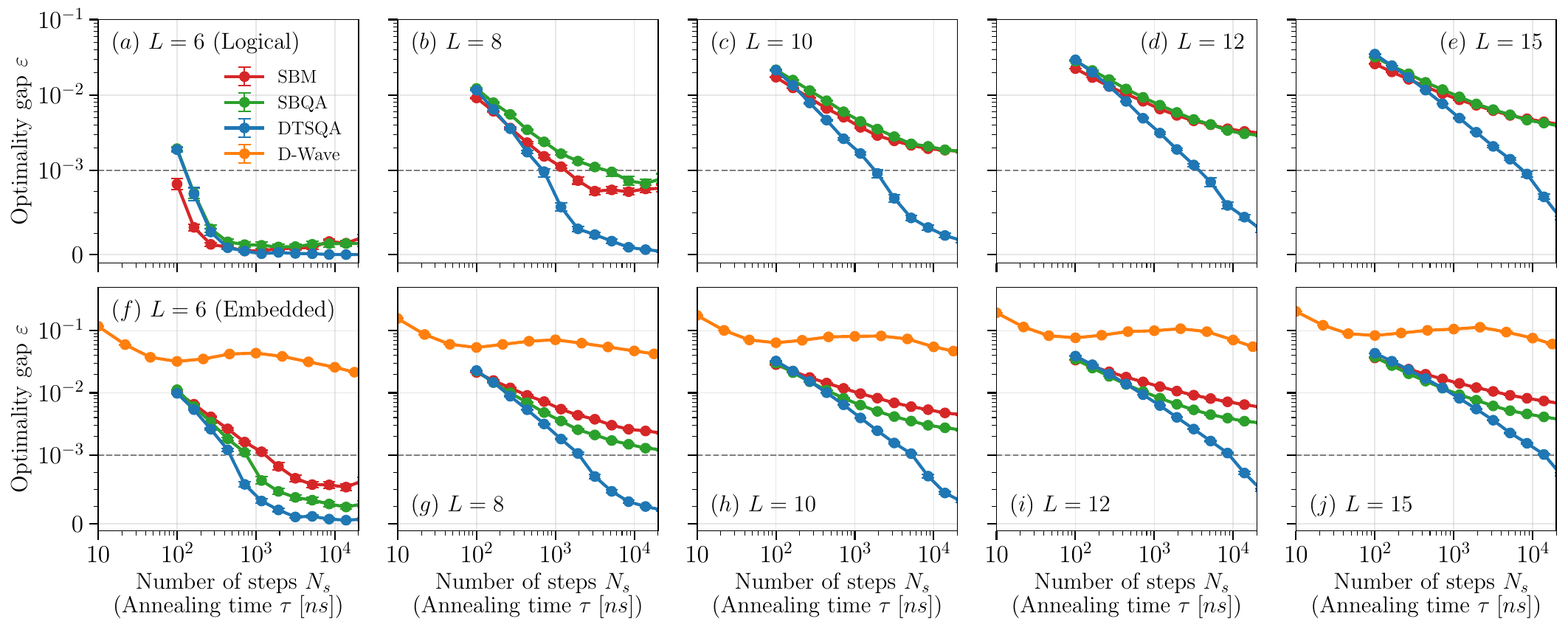}
    \caption{Optimality gap \(g\) as a function of the number of steps (annealing time in nanoseconds for D-Wave) for 
    3D spin glass instances of various sizes \(N = L^3\) for $L=6,8,10,12$ and \(N = L^2 \times 12\) for \(L=15\). Below \(\rho_E = 10^{-3}\) the y-axis
    scale switches to linear, since in some cases the algorithms are able to reach the estimated reference energy. Top row corresponds to logical instances,
    while the bottom row corresponds to instances embedded into the Pegasus \(P_{16}\) graph. We can observe that SBQA consistently achieves better solution quality than SBM for a fixed number of steps,
    except for the smallest logical instances, where the performance is slightly worse. DTSQA can provide solutions of the best quality, for a fixed number of steps.}
    \label{fig:3d_spin_glass_steps}
\end{figure}
Here, we analyze the two components of the time-to-epsilon benchmark for 3D spin glasses, namely the solution quality,
measured by the obtained optimality gap \(g\), and the runtime of the algorithms. Since the true ground state energies of these instances
are unknown, we estimate them by running DTSQA with a very large number of steps, $N_s = 5 \times 10^5$, and take the best energy found across
\(5\) runs as the reference energy \(E_0\). Crucially, the number of steps used for GS estimation is significantly larger than the maximum number of steps
used in the actual benchmarks.

\begin{figure}[htbp]
    \centering
    \includegraphics[width=\linewidth]{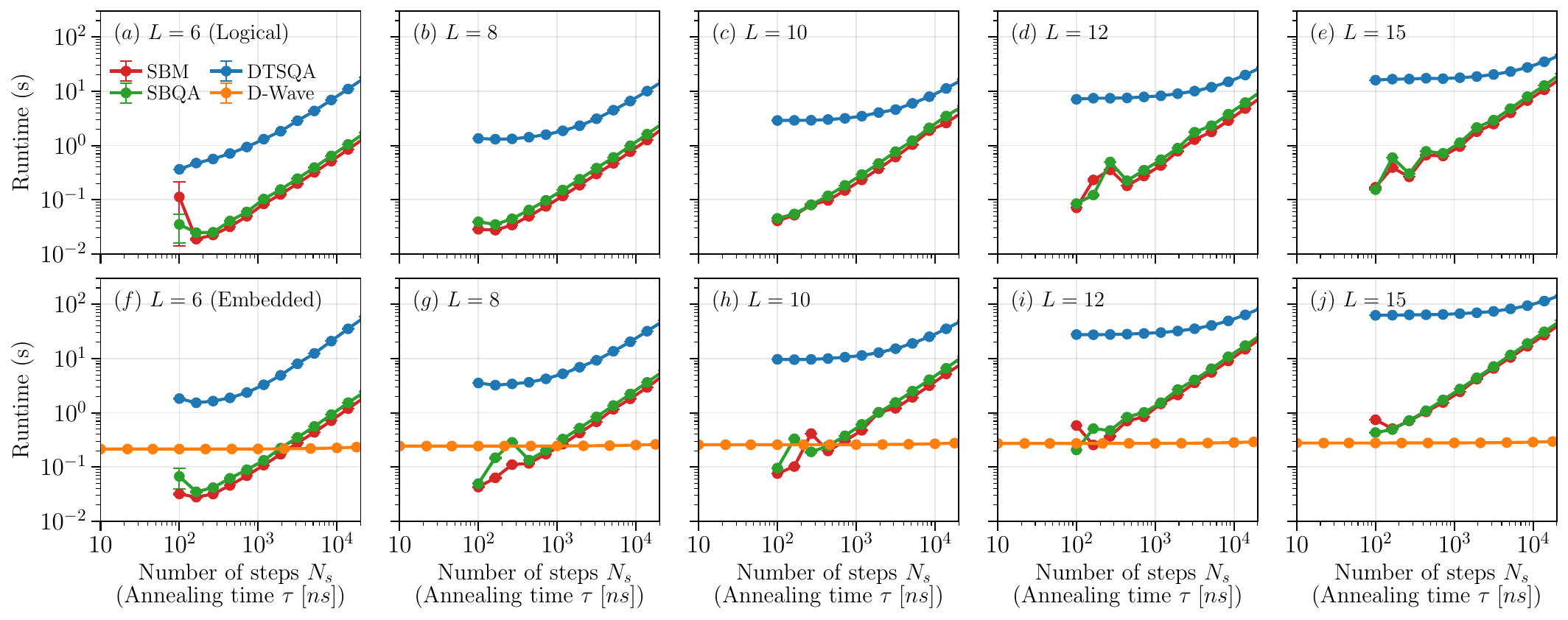}
    \caption{Measured runtime as a function of the number of steps (annealing time in nanoseconds for D-Wave) for 
    3D spin glass instances of various sizes \(N = L^3\) for $L=6,8,10,12$ and \(N = L^2 \times 12\) for \(L=15\). Top row corresponds to logical instances,
    while the bottom row corresponds to instances embedded into the Pegasus \(P_{16}\) graph. We can see that
    the runtime of SBQA is only slightly higher than that of SBM, while DTSQA is significantly slower, mainly due to the overhead associated with
    setting up the necessary data structures.}
    \label{fig:3d_spin_glass_runtime}
\end{figure}

In Fig.~\ref{fig:3d_spin_glass_steps} we plot
the optimality gap as a function of the number of steps (in the case of D-Wave annealer, the annealing time in nanoseconds), with each panel corresponding to
a fixed linear lattice size \(L\) (number of variables is O(\(L^3\))). In order to reach such small annealing times, the Advantage 4.1 QPU was operated in the fast annealing mode.
Then, in Fig.~\ref{fig:3d_spin_glass_runtime} we show how the measured runtime depends on the number of steps/annealing time for each method. 
One can see that SBQA consistently achieves better solution quality than SBM for a fixed number of steps, except for the smallest logical instances, where the performance is slightly worse.
At the same time, the runtime of SBQA is only marginally higher than that of SBM, due to the additional overhead associated with inter-replica coupling calculations.
Finally, DTSQA can provide solutions of even better quality (for a fixed number of steps), but it does so at the cost of significantly longer runtimes, which ultimately results in higher
time-to-epsilon values for moderate values of \(\varepsilon\).

\section{Additional details on HUBO benchmark\label{sec:hubo_supplemental}}
\begin{table}[htbp]
\centering
\caption{Optimal penalties by coupling distribution, $N_{\rm swap}$, and solver}
\begin{tabular}{cccccc}
\toprule
Distribution & $N_{\rm swap}$ & Penalty (SBM) & Penalty (SBQA) & Penalty (SA) & Penalty (DTSQA) \\
\midrule
Cauchy & 1 & 20.0 & 20.0 & 10.0 & 30.0 \\
Cauchy & 3 & 30.0 & 40.0 & 20.0 & 30.0 \\
Cauchy & 6 & 40.0 & 75.0 & 30.0 & 30.0 \\
Cauchy & 9 & 40.0 & 75.0 & 40.0 & 30.0 \\
\midrule
Normal & 1 & 4.0 & 6.0 & 4.0 & 10.0 \\
Normal & 3 & 6.0 & 8.0 & 6.0 & 8.0 \\
Normal & 6 & 8.0 & 8.0 & 6.0 & 6.0 \\
Normal & 9 & 6.0 & 8.0 & 6.0 & 6.0 \\
\midrule
Sym. Pareto & 1 & 10.0 & 10.0 & 8.0 & 10.0 \\
Sym. Pareto & 3 & 10.0 & 20.0 & 10.0 & 10.0 \\
Sym. Pareto & 6 & 20.0 & 30.0 & 10.0 & 8.0 \\
Sym. Pareto & 9 & 20.0 & 20.0 & 20.0 & 10.0 \\
\bottomrule
\end{tabular}
\label{tab:optimal_penalties_comparison}
\end{table}

The HUBO reduction procedure has one free parameter, the penalty coefficient, which controls the strength of the penalty terms
introduced to enforce consistency between original and auxiliary variables. Its value must be large enough for the correct reproduction
of the low-energy spectrum of the original HUBO problem. However, different solvers may benefit from different penalty values,
due to their distinct ways of exploring the solution space. Thus, for each coupling distribution and value of \(N_{\rm swap}\),
we perform a line search over penalty values for each solver, and select the value which, on average, yields the best solution quality.
This procedure ensures a fair comparison between different solvers used in our benchmarks. The optimal penalty values found
are summarized in Table~\ref{tab:optimal_penalties_comparison}.

\begin{figure}[htbp]
    \centering
    \includegraphics[width=\linewidth]{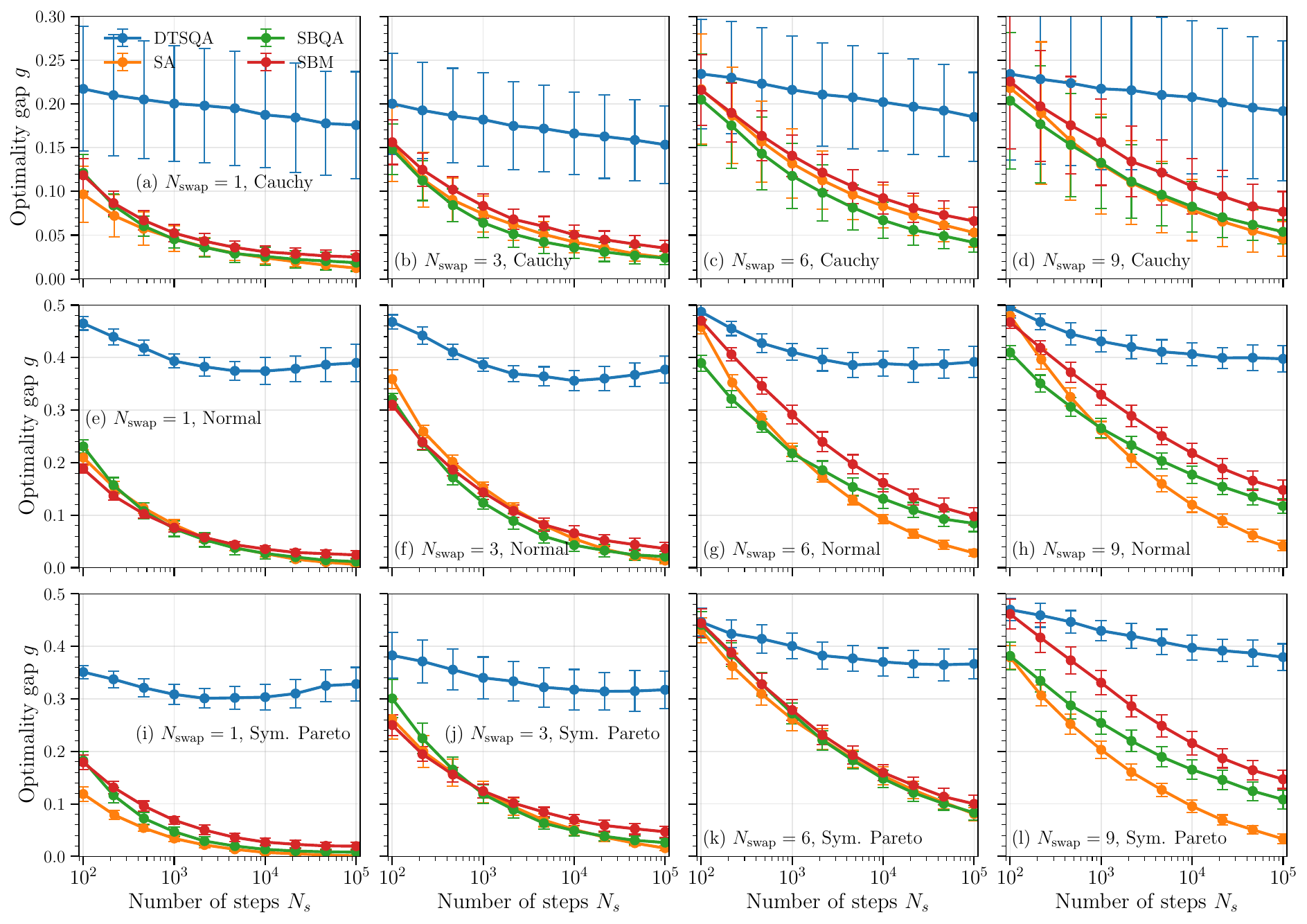}
    \caption{Optimality gap \(g\) as a function of the number of steps (sweeps for SA) for 
    QUBO-reduced HUBO instances, obtained from the heavy-hexagon topology of IBM quantum processors. Each panel
    corresponds to a fixed coupling distribution and value of \(N_{\rm swap}\) (number of SWAP operations used in the construction of the instance).
    }
    \label{fig:hubo_steps}
\end{figure}

\begin{figure}[htbp]
    \centering
    \includegraphics[width=\linewidth]{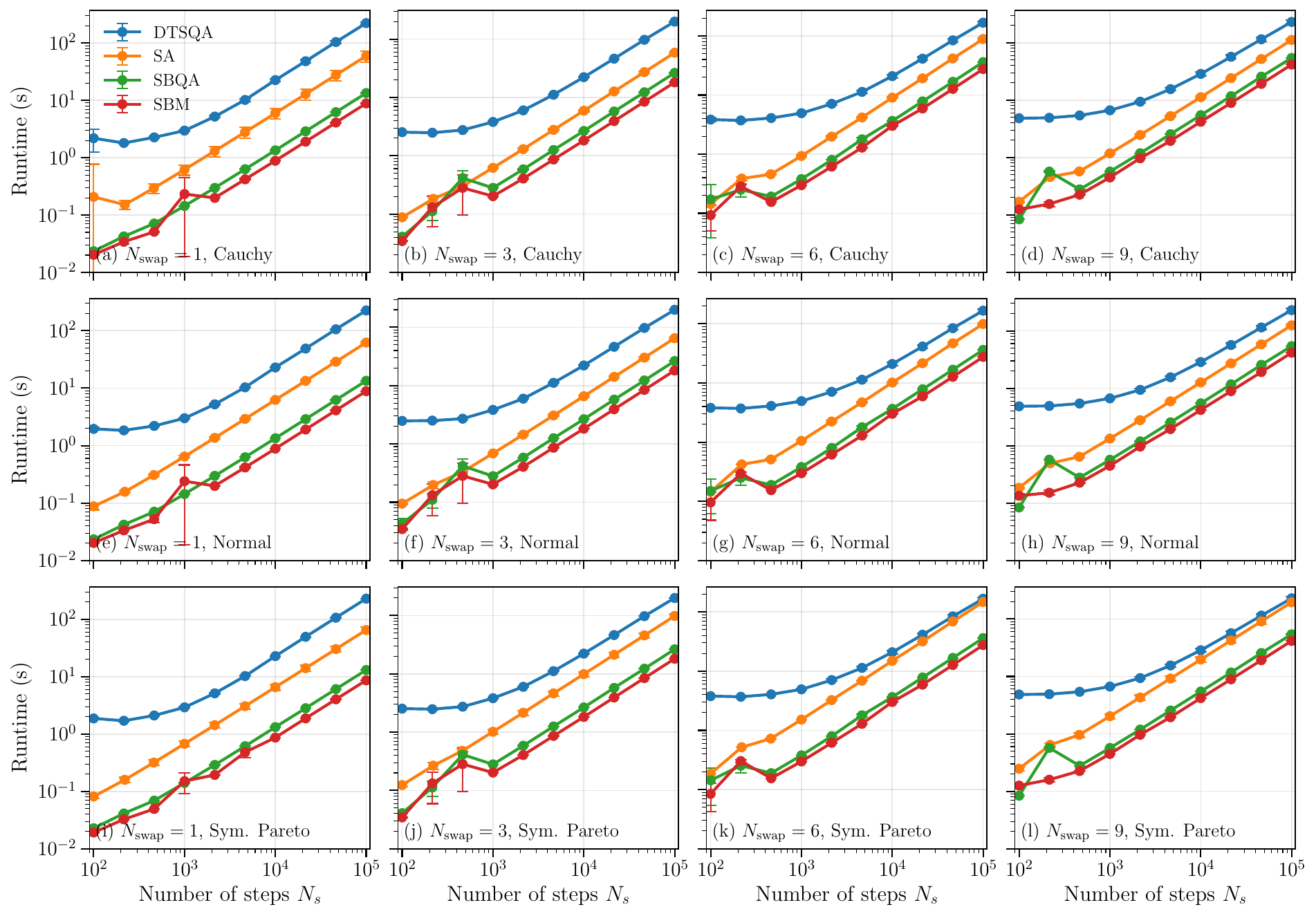}
    \caption{Measured runtime as a function of the number of steps (sweeps for SA) for 
    QUBO-reduced HUBO instances, obtained from the heavy-hexagon topology of IBM quantum processors. Each panel
    corresponds to a fixed coupling distribution and value of \(N_{\rm swap}\) (number of SWAP operations used in the construction of the instance).
    }
    \label{fig:hubo_runtime}
\end{figure}

Finally, similar to previous appendices, in Figs.~\ref{fig:hubo_steps} and~\ref{fig:hubo_runtime} we separately analyze the two components
of the time-to-epsilon benchmark for HUBO instances: solution quality and algorithm runtime. These plots clearly show how DTSQA
fails to reach solutions of similar quality as other methods, which ultimately results in infinite time-to-epsilon values for stricter
optimization targets. On the other hand, SBQA is able to consistently improve upon SBM's performance for a fixed number of steps, while maintaining
only a marginal increase in runtime. This further supports the conclusions drawn in the main text,
regarding the versatility of SBQA across a wide range of problem types and topologies.

\end{document}